\documentclass[fleqn,usenatbib]{mnras}
\usepackage{newtxtext,newtxmath,times,verbatim}
\usepackage[T1]{fontenc}
\DeclareRobustCommand{\VAN}[3]{#2}
\let\VANthebibliography\thebibliography
\def\thebibliography{\DeclareRobustCommand{\VAN}[3]{##3}\VANthebibliography}
\usepackage{graphicx}
\usepackage{amsmath}
\usepackage{ulem}
\usepackage{comment}
\usepackage{booktabs}

\title[Identifying AGN in dwarfs via variability]{Variability-selected AGN in dwarf galaxies: the incidence of AGN in dwarf and massive galaxies is similar}

\author[S. Kaviraj et al.]{S. Kaviraj,$^{1}$\thanks{E-mail: s.kaviraj@herts.ac.uk} D. De Cicco,$^{2}$ I. Lazar,$^{1}$ B. Bichang'a,$^{1}$ A. E. Watkins,$^{1}$ G. Martin,$^{3}$ \newauthor S. Koudmani$^{1,4}$\\
$^{1}$Centre for Astrophysics Research, University of Hertfordshire, Hatfield, AL10 9AB, UK\\
$^{2}$Department of Physics, University of Napoli “Federico II”, Via Cinthia 9, 80126 Napoli, Italy\\
$^{3}$School of Physics and Astronomy, University of Nottingham, University Park, Nottingham NG7 2RD, UK\\
$^{4}$St Catharine's College, University of Cambridge, Trumpington Street, Cambridge CB2 1RL, UK}

\begin{document}
\label{firstpage}
\pagerange{\pageref{firstpage}--\pageref{lastpage}}
\maketitle

\begin{abstract}
We use the VST-COSMOS survey to identify, via their optical broadband variability, 30 AGN in nearby ($z<0.4$) dwarf (10$^{8}$ M$_{\odot}$ < M$_{\rm{\star}}$ < 10$^{10}$ M$_{\odot}$) galaxies. VST-COSMOS offers a 1 deg$^2$ survey footprint, a single visit depth of 24.6 mag and 68 $r-$band visits spanning an eleven-year temporal baseline. Compared to a control sample matched in stellar mass and redshift, the dwarf AGN population shows an elevated fraction of early-type galaxies but a similar fraction of interacting objects, suggesting that interactions do not play a significant role in triggering these AGN. Dwarf AGN hosts do not show strong differences in their distances to nodes, filaments and massive galaxies compared to the controls, which indicates that AGN triggering, at least in this sample, is not strongly correlated with environment. Finally, by combining the true number of galaxies, the detectability of AGN and the measured numbers of variable sources in dwarf and massive (M$_{\rm{\star}}$ > 10$^{10}$ M$_{\odot}$) galaxies, we estimate the relative frequency of AGN in these two mass regimes. Our results suggest that the incidence of AGN in dwarfs and massive galaxies is similar (within less than a factor of 2 of each other), with some evidence that the AGN fraction increases with stellar mass in the dwarf population. 
\end{abstract}


\begin{keywords}
galaxies: evolution -- galaxies: formation -- galaxies: interactions -- galaxies:dwarf -- galaxies: active
\end{keywords}


\section{Introduction}

Given their dominance of the galaxy number density \citep[e.g.][]{Wright2017,Martin2019}, dwarf galaxies are vital for a complete understanding of galaxy evolution. {\color{black}Dwarfs have been studied in detail largely in our immediate vicinity, e.g. in the Local Group or around nearby massive galaxies \citep[e.g.][]{Tolstoy2009,Duc2015,Geha2017,Venhola2018,Trujillo2021}, with the advent of JWST \citep{Gardner2006} gradually opening up this domain in the high-redshift Universe via its near-infrared capabilities \citep[e.g.][]{Endsley2024,Witten2025,Baker2025}.} 


In the context of galaxy evolution, active galactic nuclei (AGN) are thought to significantly influence the formation of massive galaxies \citep[e.g.][]{Fabian2012,Beckmann2017}. Virtually all massive galaxies are thought to host super-massive black holes (BHs) at their centres \citep[e.g.][]{Kormendy1995,Richstone1998}. The negative feedback imparted by these BHs, when they are accreting and active, is commonly employed to regulate star formation activity in theoretical models, in order to bring predicted galaxy properties (e.g. stellar masses, morphologies and colours) in line with observational data \citep[e.g.][]{Croton2006,Kaviraj2017}. However, while BHs and AGN form an integral part of the currently-accepted picture of massive-galaxy evolution, many unexplored questions remain about BH-galaxy co-evolution in the dwarf regime. For example, what are the AGN and BH occupation fractions in dwarf galaxies? What is the relative frequency of AGN in dwarfs compared to that in their massive counterparts? Do dwarf AGN hosts show correlations with galaxy morphology and/or local environment, as is observed in the massive-galaxy regime? Is there evidence that AGN feedback regulates star formation activity in dwarfs? 

From a theoretical standpoint, BH growth in dwarfs in cosmological simulations appears to be heavily stunted. This is driven both by supernova feedback starving the BH of fuel \citep[e.g.][]{Dubois2015,Habouzit2017,Angles-Alcazar2017,Trebitsch2018,Dubois2021} and because, if not tethered to the barycentre, the shallow dwarf potential wells allow the BHs in some simulations to wander outside the central regions of the host galaxies. This results in the BHs spending most of their lifetimes in regions of low gas density, often at large distances from the central regions of the host galaxies \citep[e.g.][]{Bellovary2021,Sharma2022}. A consequence of this is that the dwarf AGN fractions in such simulations are very low (if not close to zero). Nevertheless, high-resolution idealized simulations are increasingly demonstrating the likelihood of AGN activity in simulated dwarf galaxies and the potential for AGN feedback to shape their evolution \citep{Barai2019,Koudmani2019,Sharma2023,Koudmani2024,ArjonaGalvez2024}, in a similar fashion to what is seen in massive galaxies. 

The behaviour seen in current simulations that employ cosmological volumes appears increasingly in tension with observational work. A growing literature is identifying AGN in nearby dwarf galaxies via an array of methods, e.g. using optical emission-line ratios \citep[e.g.][]{Greene2007,Reines2013,Moran2014,Dickey2019,Mezcua2024,Pucha2025}, broadband variability \citep[e.g.][]{Baldassare2020b,Burke2022,Burke2024}, infrared photometry \citep[e.g.][]{Jarrett2011,Marleau2017,Satyapal2018,Kaviraj2019}, SED-fitting \citep[e.g.][]{Zou2023,Bichanga2024}, X-ray emission \citep{Pardo2016,Chen2017,Mezcua2018,Birchall2020,Sacchi2024} and excess radio emission that cannot be accounted for by star formation alone \citep[e.g.][]{Nyland2017,Mezcua2019,Davis2022}. 

Several of these studies report appreciable fractions of AGN in the dwarf regime. {\color{black}For example, \citet{Mezcua2024} use integral field spectroscopy, from the MaNGA survey \citep{Bundy2015}, of dwarf galaxies drawn from the SDSS to show that at least 20 per cent (and up to $\sim$54 per cent) of these systems show signs of AGN activity in their optical emission line ratios. Notably off-nuclear AGN, which are difficult to detect using single fibre spectroscopy, significantly contribute to the statistics in this study.} \citet{Dickey2019} use a similar methodology, implemented via long-slit spectroscopy, to show that $\sim$80 per cent of quiescent field dwarfs in the local Universe show AGN signatures in their central regions. SED fitting using deep UV to mid-infrared broadband photometry indicates that around a third of dwarfs show signs of  AGN activity \citep{Bichanga2024}, while \citet{Davis2022}, who combine deep radio and optical data to select radiatively-inefficient AGN, conclude that AGN triggering in dwarfs is likely to be stochastic and a common phenomenon. 

The existence of AGN in nearby dwarfs appears consistent with recent observations using the JWST which reveal the widespread presence of AGN in low-mass galaxies at high redshift \citep[e.g][]{Scholtz2023}, and the conclusions of some studies that AGN activity appears common in such systems \citep[e.g.][]{Juodzbalis2023}. {\color{black}While nearby dwarfs are not descendants of galaxies that have similar masses at high redshift, the JWST results demonstrate that AGN activity can routinely take place within the gravitational potential wells that host low-mass galaxies (see \citealt{Mezcua2024b} for a discussion of the link between AGN discovered in the high redshift Universe via JWST and those that reside in nearby dwarf galaxies).} Taken together, these empirical results spanning a range of redshifts raise the possibility that AGN activity (and therefore BH growth) may have existed in the dwarf regime over cosmic time. This aligns with the fact that nearby dwarfs appear to lie on an extrapolation of the $M$ -- $\sigma$ relation defined by massive galaxies \citep[e.g.][]{Schutte2019,Davis2020,Baldassare2020a}, a correlation that is expected to arise naturally if AGN activity regulates the growth of its host galaxy \citep[e.g.][]{Silk1998,King2021}.  

As we explore in detail in Section \ref{sec:frequency}, while many AGN have indeed been discovered in dwarfs, accurate measurements of the AGN fraction remain challenging because AGN detectability in the dwarf regime is often low. For example, typical dwarfs outside the local neighbourhood are too faint to be detectable in past wide-area surveys like the SDSS because they are shallow \citep{Kaviraj2025}. The dwarfs that are detected in shallow surveys outside the very local Universe are biased towards star-forming galaxies, in which young stars boost the galaxy luminosities above the survey detection thresholds of shallow datasets like the SDSS. {\color{black} Unbiased statistical studies of dwarfs require surveys that are both deep and wide. The depth is required for detecting typical dwarfs outside the local neighbourhood and large areas are needed to create statistical samples of dwarfs, particularly in low-density environments. While this will become routinely possible using new and forthcoming surveys like the Legacy Survey of Space and Time \citep[LSST;][]{Ivezic2019} and Euclid \citep{Laureijs2011}, some precursor datasets already exist (such as in the COSMOS field) where such studies are possible, albeit with smaller number of galaxies.}

The use of shallow surveys also complicates the detection of AGN activity in dwarf galaxies, since it can be swamped by the high levels of star formation that exist in the dwarfs that are present in these datasets  \citep[e.g.][]{Mezcua2024}. In a similar vein, the depth of any ancillary data (e.g. radio fluxes) that is being used to identify AGN adds an extra layer of complexity to the detection process and can further reduce the detectability of AGN in dwarf galaxies. A consequence of these two issues, which is particularly important in the dwarf regime where both the host galaxy and the AGN can be relatively faint, is that an accurate measurement of the AGN fraction demands knowledge of both host and AGN detectability. Measuring detectability, however, is challenging, as a result of which this quantity is typically ignored in most studies. We explore this issue further, in the context of the data used in this study, to derive an estimate of the relative frequency of AGN in the dwarf and massive-galaxy regimes, in Section \ref{sec:frequency} below.  

{\color{black}Given the biases noted above, techniques such as variability \citep[e.g.][]{Elmer2020,Kimura2020,Burke2024}, that can identify AGN in dwarfs using (deep) broadband photometry alone, are desirable.} Deep photometry makes it possible to detect faint galaxies outside the local neighbourhood, and a survey which is both deep and wide can then be used to construct statistically meaningful samples of dwarfs outside the local Universe. Indeed, the shape of the galaxy mass function means that even modest survey areas can yield large samples of dwarfs. The use of broadband optical variability holds particular promise in the near future with the advent of the Legacy Survey of Space and Time \citep[LSST; ][]{Ivezic2019,Watkins2024}, which will offer a three-day cadence, a single-visit depth of $r\sim24.7$ magnitudes, an 18,000 deg$^2$ footprint and, at completion, a ten-year temporal baseline. 

In this paper, we use optical broadband variability to identify AGN in nearby ($z<0.4$) dwarf (10$^{8}$ M$_{\odot}$ < M$_{\rm{\star}}$ < 10$^{10}$ M$_{\odot}$) galaxies, using the VST-COSMOS survey. This survey offers a virtually identical single-visit depth and temporal baseline to what will be available from the LSST at completion. Notwithstanding its 1 deg$^2$ area, VST-COSMOS enables us to extract statistically significant samples of nearby dwarfs and identify AGN within them via variability, offering a preview of the game-changing science that will be possible using the LSST in the near future. 

The plan for this paper is as follows. In Section \ref{sec:data}, we describe the selection of dwarf galaxies that host AGN using VST-COSMOS, the morphological classification of these dwarfs via visual inspection and the estimation of environmental parameters (distances to the nearest node, filament and massive galaxy) using the DisPerSE algorithm \citep{Sousbie2011}. In Section \ref{sec:properties}, we study the properties of our AGN and compare them to the general dwarf population, using control samples matched in stellar mass and redshift. We explore the fraction of the optical flux that might be contributed by the AGN, the role of interactions in AGN triggering and the morphologies, star formation rates (SFRs) and environments of our AGN hosts compared to the general dwarf population. In Section \ref{sec:frequency}, we use the VST-COSMOS data to quantify the relative frequency of AGN in dwarfs compared to that in massive (M$_{\rm{\star}}$ > 10$^{10}$ M$_{\odot}$) galaxies. We summarise our findings in Section \ref{sec:summary}.


\section{Data}

\label{sec:data}

 
\subsection{A catalogue of AGN selected via variability from the VST-COSMOS survey}

The sample of AGN in dwarf galaxies that underpins this study is constructed from observations of the COSMOS field using OmegaCAM \citep{Kuijken2011} on the VST \citep{Capaccioli2011} at the Cerro Paranal observatory. The construction of the AGN catalogue follows the procedure described in \citet{Decicco2015} and \citet{Decicco2019} and is briefly outlined here. Lightcurves are calculated using VST observations for an initial sample of 22,927 sources defined in \citet{Decicco2019}. The VST-COSMOS dataset used here consists of 68 $r-$band observations spanning eleven years. A variability threshold is computed by calculating the average magnitude over the eleven-year baseline and the corresponding root mean square (RMS) deviation of each source from its lightcurve. The sample of variable sources is then (conservatively) defined as those in which the RMS deviations are in excess of the 95th percentile of the distribution of these values. In order to identify spurious sources, the snapshots and lightcurves of each variable source are visually inspected and each source is flagged with a quality label from 1 to 3. Sources with a flag of 1 are strong candidates without any obvious problems, those with a flag of 2 are very likely to be variable but have a neighbour, while those with a flag of 3 could be spurious, mostly because of the presence of a very close neighbour. Although the single-visit depth is of 24.6 mag, we limit our analysis to sources brighter than 23.5 mag, in order to minimize contamination from noisy objects. Following  \citet{Decicco2015} and \citet{Decicco2019}, a `robust' sample is then defined using sources with a flag of 1 and 2 only. 

{\color{black}We briefly discuss other potential sources of galaxy variability -- e.g. tidal disruption events (TDEs), supernovae and host variability due to fluctuations in the star formation history -- and describe why they do not contaminate the AGN sample used here. TDEs exhibit well-defined peaks and decay smoothly over time \citep[e.g.][]{Zabludoff2021}. In addition, they are also initially much more luminous than what is seen in AGN, with the object brightening by several magnitudes. However, objects in our variable sample exhibit stochastic variability over an eleven-year timescale and, as discussed in Section \ref{sec:flux_fractions}, the root mean square (RMS) of the variable magnitude is less than 0.15 mag. In a similar vein, supernovae are also ruled out because their lightcurve shapes, which involve a rapid rise followed by a slower decay, are qualitatively different to those of AGN. For example, in \citet{Decicco2015}, which employs the same AGN selection procedure as in this work and has a 5 month baseline, the rate of contamination by supernovae is $\sim$10 per cent. However, in \citet{Decicco2019}, which utilizes a 3.3 year baseline (significantly shorter than the baseline in this study) the supernova contamination rate drops to zero. 

Finally, the variability is very unlikely to be driven by the star formation history (SFH) of the galaxies due to two reasons. First, the dynamical timescales of dwarf galaxies, over which we would expect to observe variability driven by the star formation rate (SFR), are hundreds of Myrs (orders of magnitude longer than our eleven-year baseline). Second, if the variability was driven by fluctuations in the SFR then we would expect a correlation between variability and colour and, in particular, see frequent (and preferential) variability in blue galaxies. However, the blue fraction is significant (around 50 per cent, see e.g. \citealt{Kaviraj2025}) and much larger than the fraction of variable objects, making it unlikely that the variability is driven by host galaxy fluctuations due to changes in the SFR.}


\subsection{A catalogue of dwarf galaxies that host AGN}
\label{sec:dwarfagn}

To construct our catalogue of dwarf AGN, we combine the variable sample described above with physical parameters (photometric redshifts, stellar masses, rest-frame colours and SFRs) from the Classic version of the COSMOS2020 catalogue \citep{Weaver2022}. The physical parameters in this dataset are calculated by applying the \textsc{LePhare} SED-fitting algorithm \citep{Arnouts2002,Ilbert2006} to deep, multi-wavelength UV to mid-infrared photometry in around 40 broad and medium band filters, from the following instruments: GALEX \citep{Zamojski2007}, MegaCam/CFHT \citep{Sawicki2019}, ACS/HST \citep{Leauthaud2007}, Hyper Suprime-Cam \citep{Aihara2019}, Subaru/Suprime-Cam \citep{Taniguchi2007,Taniguchi2015}, VIRCAM/VISTA \citep{McCracken2012} and IRAC/Spitzer \citep{Ashby2013,Steinhardt2014,Ashby2015,Ashby2018}. 

A particular novelty is the inclusion of optical ($i,z$) data from the ultra-deep layer of the Hyper Suprime-Cam Subaru Strategic Program (HSC-SSP) for object detection, which has a point-source depth of $\sim$28 mag \citep{Aihara2019}. As a comparison, this is $\sim$10 mag deeper than the magnitude limit of the SDSS spectroscopic main galaxy sample \citep[MGS; e.g.][]{Alam2015}. The accuracy of the  photometric redshifts in COSMOS2020 is better than 1 and 4 per cent for bright ($i<22.5$ mag) and faint ($25<i<27$ mag) galaxies respectively. 

To construct the sample of dwarf AGN that underpins this study we proceed as follows. We first select objects which are both classified as galaxies by \textsc{LePhare}  (`type' = 0 in the COSMOS2020 catalogue) and as `extended' (i.e. galaxies) in the HSC $griz$ filters. We then select galaxies which have stellar masses and redshifts in the ranges 10$^{8}$ M$_{\odot}$ < M$_{\rm{\star}}$ < 10$^{10}$ M$_{\odot}$ and $z<0.4$ respectively, which lie within the HSC-SSP footprint and outside bright-star masks and which have both $u$-band and mid-infrared photometry (since a long wavelength baseline improves the accuracy of the parameter estimation, e.g. \citet{Ilbert2006}). The choice of redshift and stellar mass ranges is driven by the fact that, as described in Section \ref{sec:properties}, the galaxy population within these ranges is complete. This parent dwarf sample is cross matched with the sample of variable systems, producing 30 dwarf AGN within the stellar mass and redshift ranges described above. For the purposes of Section \ref{sec:frequency}, we also construct a sample of massive galaxies (which are defined throughout this study as those with M$_{\rm{\star}}$ > 10$^{10}$ M$_{\odot}$) in an identical fashion and cross-match it with the variable sample to yield 10 massive galaxies which host AGN at $z<0.4$. The AGN in massive galaxies will be studied in detail in a forthcoming paper (Bichang'a et al. in preparation). 

{\color{black}The SED fitting used to derive the physical parameters we use here does not incorporate AGN templates. It is worth noting in this context, however, that current AGN templates are largely built and calibrated based on luminous AGN in massive galaxies and are unlikely to be appropriate for dwarf galaxies. Here, we do not consider X-ray sources, which could host powerful AGN (which might potentially affect the SED fitting)\footnote{{\color{black}Note that \citet{Burke2024} have also recently studied optically variable AGN in the COSMOS field. Their sample is based on the catalogue produced} {\color{black}by \citet{Kimura2020}, who derive optically variable galaxies from the HSC-SSP using a three-year baseline. However, around 90 per cent of the Kimura et al. sample has X-ray detections, which rules them out of our selection criteria. Furthermore, the Burke et al. study focuses on massive galaxies (see, for example, their Figure 10) and therefore does not overlap with the stellar mass regime studied in this paper.}}. Appendix \ref{app:pzsz} shows a comparison between the COSMOS2020 photometric redshifts and spectroscopic redshifts of 16 variable dwarfs which have available spectroscopic redshifts, from the recent compilation of \citet{Khostovan2025}. The photometric redshifts show excellent correspondence with their spectroscopic counterparts, with a median difference between the photometric and spectroscopic redshifts of $\sim$0.0066. This suggests that, in spite of the presence of an AGN, the SED fitting produces reliable physical parameters in our sample of galaxies. Note that the upper mass limit of the dwarf regime can vary in literature, with slightly lower limits adopted by some studies \citep[e.g.][]{Vandermarel2002,Davis2022}. Together with the fact that stellar mass errors from SED fitting can be of the order of a few tenths of a dex \citep[e.g.][]{Weaver2022} and that the presence of AGN may lead to overestimates of stellar mass \citep[e.g.][]{Siudek2024}, our choice of upper mass limit for the dwarf regime is reasonable.}


\subsection{Morphological classifications using visual inspection}

\begin{figure*}
\center
\includegraphics[width=2\columnwidth]{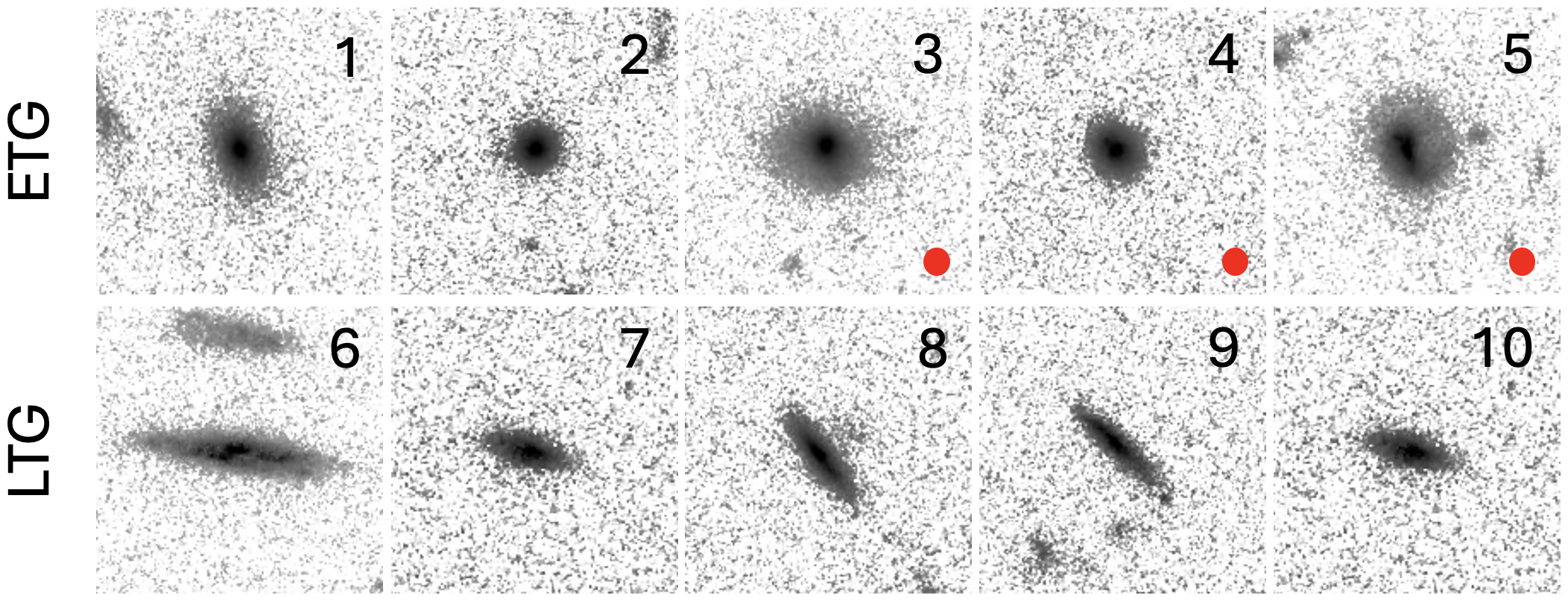}
\caption{Examples images of our dwarf AGN (each image has a linear size of 4 arcsec). The top and bottom rows show early-type galaxies (ETGs) and late-type galaxies (LTGs) respectively. Interacting systems are indicated using a red filled circle in the lower right-hand corner of the image. Galaxies 3 and 5 show internal asymmetries, while galaxy 4 appears to be accreting a smaller companion in its central region. {\color{black}This figure is better viewed online than in print.}} 
\label{fig:images}
\end{figure*}

We use visual inspection of Hubble Space Telescope (HST) F814W ($I$-band) images in the COSMOS field \citep{Koekemoer2007,Massey2010} to morphologically classify our dwarf AGN and control samples (described in Section \ref{sec:control} below). We classify our dwarfs by eye into two broad morphological classes: early-type galaxies (ETGs) and late-type galaxies (LTGs). We also identify galaxies that show evidence of a current or recent interaction, such as internal asymmetries or an ongoing merger with another galaxy. Figure \ref{fig:images} presents example images of our ETGs (top row) and LTGs (bottom row). Interacting systems are indicated using a red filled circle in the lower right-hand corner. Galaxies 3 and 5 show internal asymmetries, while galaxy 4 appears to be accreting a smaller companion in its central regions. 

{\color{black}We note that an array of methods have been used in the literature to classify galaxy morphologies, ranging from visual inspection of images to algorithmic techniques which employ morphological parameters. However, visual inspection is typically considered the most accurate technique of morphological classification \citep[e.g.][]{Kaviraj2014b,Lintott2011}, against which other commonly used methods like morphological parameters \citep[e.g.][]{Conselice2003,Lotz2004} are calibrated. While morphological parameters are typically employed for large datasets for which visual inspection can be prohibitively time consuming, the size of our dataset lends itself well to morphological classification via this method. More importantly, recent work has shown that classical parameters like the CAS system, M$_{20}$ and the Gini coefficient are not effective at separating dwarfs galaxies into ETGs and LTGs \citep{Lazar2024a}. It is also worth noting that, unlike their massive counterparts, dwarf ETGs and LTGs overlap significantly in their rest-frame optical colour distributions \citep{Lazar2024b}, as a result of which colour information cannot be reliably used to separate morphological types in this regime.} 

\subsection{Environmental parameters calculated using DiSPerSE}
\label{sec:environment} 

We use DisPerSE, a structure-finding algorithm, to measure the locations of galaxies within the cosmic web, such as their distances to the nearest nodes, filaments and massive  galaxies. The algorithm measures the density field via Delaunay tessellations, which are calculated using the positions of galaxies \citep{Schaap2000}. It then uses this density field to identify the positions of critical points i.e. minima, maxima and saddles, which correspond to the locations of voids, nodes and the centres of filaments respectively. Segments are then used to connect pairs of saddle points and local maxima, forming a set of ridges that describe the network of filaments that define the cosmic web. 

A `persistence' parameter sets a threshold value for retaining pairs of critical points within the density map. For example, a persistence of `\textit{N}' results in all pairs which have Poisson probabilities below \textit{N}$\sigma$ from the mean being removed. Here, we set the persistence equal to 2, following the methodology of \citet{Laigle2018}, who have implemented DisPerSE on redshift slices of similar widths as in our analysis, constructed from the COSMOS2015 \citep{Laigle2016} catalogue. We note that the same methodology has been used to do an identical density analysis using the COSMOS2020 catalogue by \citet{Lazar2023} and \citet{Bichanga2024}. We refer readers to \citet{Sousbie2011} for further details about the algorithm. 



The accurate COSMOS2020 redshifts enable us to employ relatively narrow redshift slices to build density maps. We only use massive galaxies for this purpose, because they have the smallest redshift errors and dominate the local gravitational potential well. Individual massive galaxies are weighted by the area under their redshift probability density functions that is contained within the redshift slice in question. This takes into account the fact that the photometric redshifts in COSMOS2020, albeit very accurate, do have associated uncertainties. 

{\color{black}Finally, we consider the types of environments that are likely to be present in the VST-COSMOS footprint in our redshift range of interest, by considering the \textit{M}$_{200}$ values (which are proxies for the virial masses) of groups identified in the literature \citep{Finoguenov2007,George2011,Gozaliasl2014,Gozaliasl2019}. The virial masses of groups in VST-COSMOS lie in the range 10$^{12.9}$ M$_{\odot}$ < \textit{M}$_{\rm{200}}$ < 10$^{13.8}$ M$_{\odot}$, with a median value of 10$^{13.4}$ M$_{\odot}$. For comparison, a small cluster like Fornax has a virial mass of $\sim$10$^{13.9}$ M$_{\odot}$ \citep{Drinkwater2001}, while large clusters like Virgo and Coma have virial masses of $\sim$10$^{15}$ M$_{\odot}$ \citep[e.g.][]{Fouque2001,Gavazzi2009}. The galaxy population considered in this study therefore spans a wide spectrum of environments from large groups to the field but does not reside in rich clusters.}


\section{Properties of dwarf galaxies that host AGN} 
\label{sec:properties}

\begin{figure}
\center
\includegraphics[width=\columnwidth]{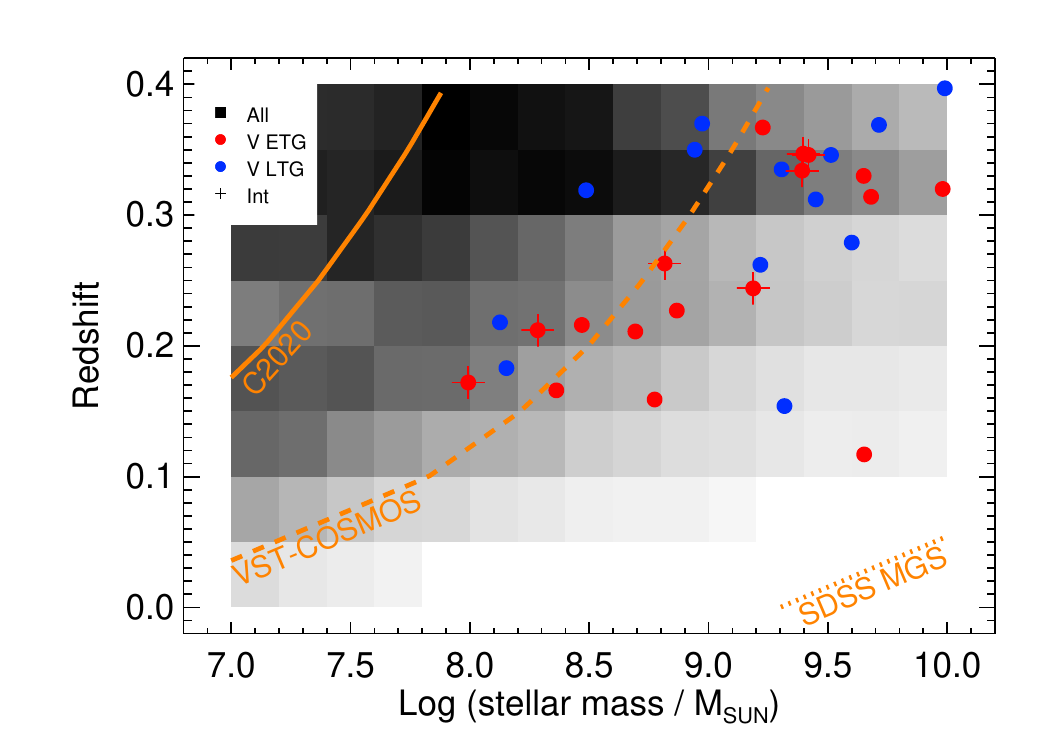}
\caption{Redshift vs stellar mass of dwarfs in the COSMOS2020 catalogue (shown using the heatmap) and our dwarf AGN (filled circles). ETGs and LTGs are shown colour-coded (see legend). Interacting galaxies are indicated using crosses. The solid, dashed and dotted orange lines show the redshifts below which galaxies of a given stellar mass are complete in COSMOS2020, VST-COSMOS and the SDSS MGS, which have magnitude limits of $r=28$, 24.6 and 17.77 respectively.} 
\label{fig:complete}
\end{figure}

In Figure \ref{fig:complete}, we present the stellar masses and redshifts of our dwarf AGN. The heatmap shows the COSMOS2020 galaxy population, while the filled circles indicate our dwarf AGN. Different morphological classes are shown colour-coded (see legend), while crosses indicate interacting galaxies. The solid orange line shows the redshift at which a galaxy population of a given stellar mass is likely to be complete in COSMOS2020, constructed using the methodology of \citet{Kaviraj2025}. This is defined as the redshift at which a purely-old `simple stellar population' (SSP) of a given stellar mass that forms in an instantaneous burst at $z=2$, is detectable, at the depth of the HSC ultra-deep imaging in COSMOS (which underpins object detection in the COSMOS2020 catalogue\footnote{Note that hardly any dwarfs are consistent with a purely old stellar population \citep[e.g.][]{Lazar2024a}, making this a conservative criterion.}). We construct our SSPs using the \citet{Bruzual2003} stellar models, assuming half solar metallicity. 

This purely-old SSP represents a faintest `limiting' case, since real galaxies, which are not composed uniquely of old stars, will be more luminous than this limiting value. Thus, if this limiting case is detectable at a given survey depth, then the entire galaxy population at a given stellar mass will also be detectable in the survey in question. The parameter space below the solid orange line therefore represents the region where the COSMOS2020 galaxy population is complete. Figure \ref{fig:complete} indicates that galaxy populations in COSMOS2020 with stellar masses greater down to 10$^{8}$ M$_{\odot}$ are complete out to $z\sim0.4$ (note that \citet{Weaver2022} come to an identical conclusion). As a comparison, the dashed and dotted orange lines show the corresponding completeness thresholds for VST-COSMOS and the SDSS MGS respectively. 

Note that this does not mean that galaxies will not exist in a given survey above the curves that define these completeness regions. Rather, galaxy populations will become progressively more biased towards bluer, star forming systems as we move above and further away from the curves. Figure \ref{fig:complete} indicates that the vast majority of our dwarf AGN also fall within the completeness region for VST-COSMOS. However, only galaxies at the upper mass end of our study would be complete in the SDSS MGS, and that too only in the very local Universe.  

\subsection{Construction of control samples}
\label{sec:control}

To enable comparisons between our dwarf AGN and the general dwarf population, we construct control samples which consist of populations of non-variable dwarfs with the same mass and redshift distribution and five times the number of galaxies as our dwarf AGN. A control population is constructed by identifying, for every AGN, all non-AGN within stellar mass and redshift tolerances of 0.05 dex and 0.03 respectively. We then select, at random, five of these galaxies to be the control counterparts of the AGN in question. No control galaxy is assigned to more than one AGN. 

Recall that, as described above, incompleteness will tend to bias the galaxy population towards bluer galaxies. In the massive-galaxy regime, bluer galaxies tend to have more late-type morphology \citep[e.g.][]{Buta1994,Strateva2001}, although this effect is much less pronounced in the dwarf regime, where ETGs and LTGs, at least in the nearby Universe ($z<0.08$), exhibit similar rest-frame colour distributions \citep{Lazar2024a}. Thus, to check that incompleteness does not bias our conclusions, we perform our AGN vs control comparisons in two ways. First, we construct a control population that is matched in redshift and stellar mass to our full sample of 30 dwarf AGN. Second, we construct a different control sample that is matched in redshift and stellar mass to the 22 dwarf AGN that reside within the VST-COSMOS completeness region i.e. below the dashed orange line in Figure \ref{fig:complete}. We then compare the properties of each control sample and its corresponding dwarf AGN population. 

Table \ref{tab:comparisons} summarises the comparisons between our dwarf AGN and control galaxies, which are described in detail in Sections \ref{sec:morphology}, \ref{sec:starformation} and \ref{sec:cosmicweb} below. The upper section of this table considers all dwarf AGN, while the lower section considers the case where our dwarf AGN are restricted to the completeness region of VST-COSMOS. It is worth noting that, within the uncertainties, the values in both sections of Table \ref{tab:comparisons} are consistent with each other. In other words, restricting our dwarf AGN to the VST-COSMOS completeness region does not alter our conclusions. This is likely driven by the fact that 73 per cent of our dwarf AGN lie within the VST-COSMOS completeness region and, those that do not, still reside close to the completeness threshold (i.e. the dashed orange line in Figure \ref{fig:complete}). 


\subsection{AGN flux fractions}
\label{sec:flux_fractions}

\begin{figure}
\center
\includegraphics[width=\columnwidth]{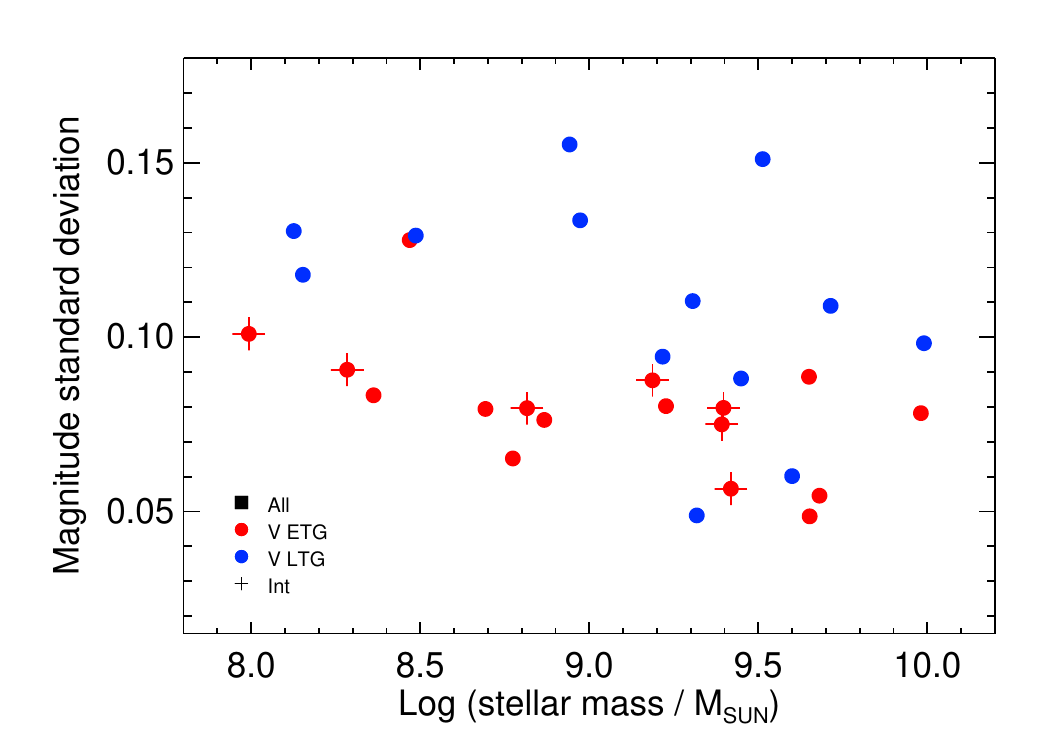}
\caption{The standard deviation of the $r$-band magnitude in our dwarf AGN. Different morphological classes are shown colour-coded (see legend). Interacting galaxies are indicated using crosses. As described in Section \ref{sec:flux_fractions}, if all the variability can be attributed to the AGN, then this standard deviation is also an estimate of the fraction of $r$-band flux that is likely to be contributed by the AGN.} 
\label{fig:flux_fractions}
\end{figure}

It is instructive to first consider the fractions of flux that may be contributed by the AGN in our variable dwarf population. Figure \ref{fig:flux_fractions} presents the standard deviation of the $r$-band magnitude in our variable dwarfs over our 11-year baseline, as a function of their stellar mass. 
This quantity traces the extent of the variability in our dwarf AGN. {\color{black}Differentiating the magnitude equation, $m=-2.5 \log (F)$ + constant, where $m$ is the magnitude and $F$ is the flux, yields $\Delta m = 1.09$ $\times$ $(\Delta F)/F$.} If the variability is attributed to the AGN, then the standard deviation ($\Delta m$) provides an estimate of the fraction of flux that is likely to be contributed by the AGN. The $r$-band flux fractions of our dwarf AGN are modest and range from $\sim$5 to $\sim$15 per cent. The median flux fraction in AGN that are hosted by LTGs is around a factor of 1.4 higher than that in their ETG counterparts. 


Finally, it is interesting to note that the flux fractions in dwarf AGN that are classified as interacting are not larger than in their non-interacting counterparts. {\color{black}If the flux fractions indicate the AGN accretion rates relative to the brightness of the host galaxy, then this suggests that interactions are unlikely to drive higher accretion rates in our dwarf AGN (see also \citealt{Erostegui2025})}. 

\subsection{Morphology and the role of interactions} 
\label{sec:morphology}

We begin our analysis of how our dwarf AGN compare to the general dwarf population by considering their morphological properties. The first row in both sections of Table \ref{tab:comparisons} indicates that the fraction of ETGs is elevated compared to that in the control sample. However, within the uncertainties, the fraction of interacting galaxies in the AGN and control populations (second row in both sections of the table) are indistinguishable, suggesting that, at least in this sample of dwarf AGN, interactions appear unlikely to play an important role in AGN triggering. {\color{black}This is similar to the findings of other studies in the recent literature \citep[e.g.][]{Erostegui2025}}. It is interesting to note that the only interacting systems that host dwarf AGN are those with early-type morphology. This may indicate that the presence of interactions makes it more likely that AGN are triggered in dwarf ETGs, while they do not similarly impact the likelihood of AGN triggering in LTGs. 
  
The comparable interacting fractions in our AGN and control samples are consistent with the findings of \citet{Bichanga2024}, who have used an identical visual classification methodology to compare the morphological properties of nearby dwarf AGN (identified via SED fitting) to that in the general dwarf population. However, unlike this study, Bichang'a et al. do not find a difference in the morphological mix of dwarf AGN and the reasons for this discrepancy are unclear. 


\begin{table}
\begin{center}
\begin{tabular}{ c | c | c }

\multicolumn{3}{c}{All dwarf AGN}\\
\toprule
& AGN & Control\\
\midrule
Early-type fraction & 0.57$^{\pm 0.09}$ & 0.39$^{\pm 0.04}$\\
Interacting fraction & 0.23$^{\pm 0.08}$ & 0.16$^{\pm 0.03}$\\
Quenched fraction & 0.14$^{\pm 0.06}$ & 0.23$^{\pm 0.03}$\\\\\\

\multicolumn{3}{c}{Dwarf AGN in completeness region only}\\
\toprule
& AGN & Control\\
\midrule
Early-type fraction & 0.64$^{\pm 0.10}$ & 0.36$^{\pm 0.05}$\\
Interacting fraction & 0.23$^{\pm 0.09}$ & 0.15$^{\pm 0.03}$\\
Quenched fraction & 0.18$^{\pm 0.08}$ & 0.32$^{\pm 0.04}$\\\\

\end{tabular}
\caption{Comparisons between our dwarf AGN and control galaxies. The upper section of this table considers all dwarf AGN, while the lower section considers the case where our dwarf AGN are restricted to the completeness region of VST-COSMOS (i.e. the region below the dashed orange line in Figure \ref{fig:complete}). In each section, we show the fractions of galaxies that are ETGs (first row), interacting (second row) and quenched (third row) in the dwarf AGN (left-hand column) and the corresponding control population (right-hand column). Uncertainties are calculated following \citet{Cameron2011} and shown using superscripts.} 
\label{tab:comparisons}
\end{center}
\end{table}


\subsection{Star formation activity} 
\label{sec:starformation}

\begin{figure}
\center
\includegraphics[width=\columnwidth]{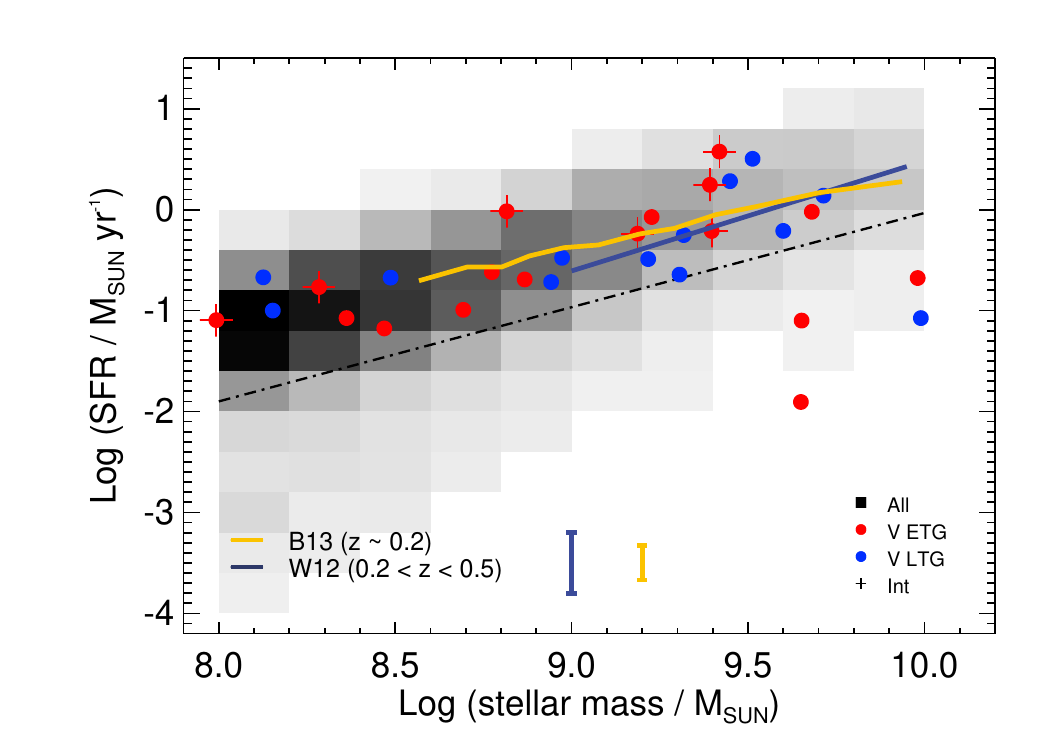}
\caption{The star formation main sequence of galaxies in the COSMOS2020 catalogue (shown using the heatmap) and our dwarf AGN (filled circles). The dashed-dotted line demarcates the lower limit of the main locus of the star formation main sequence. Different morphological classes are shown colour-coded (see legend). Interacting galaxies are indicated using crosses. Examples of other main sequence loci from \citet[][B13]{Behroozi2013} and \citet[][W12]{Whitaker2012} are shown overplotted and their standard deviations are indicated using the error bars.} 
\label{fig:sfms}
\end{figure}

We proceed by considering, in Figure \ref{fig:sfms}, the star formation rates (SFRs) of our dwarf AGN. {\color{black}Recall here that the estimation of SFRs does not include AGN templates but that the excellent correspondence between spectroscopic and photometric redshifts indicates that the physical parameters (e.g. stellar masses, redshifts and SFRs) derived via SED fitting are robust.} The heatmap shows the star formation main sequence of the COSMOS2020 galaxy population in our redshift range of interest ($z<0.4$), while the filled circles indicate our dwarf AGN. 
The dashed-dotted line indicates the lower limit of the main locus of the star formation main sequence. We consider galaxies that fall below this threshold to be quenched. Examples of main sequence loci from \citet{Behroozi2013} and \citet{Whitaker2012}, which agree well with the COSMOS2020 locus, are shown overplotted and their standard deviations are indicated using the error bars. The third row in each section of Table \ref{tab:comparisons} compares the quenched fractions in the dwarf AGN and control samples. 

Taking the uncertainties into account, the quenched fraction in dwarf AGN does not show differences with that in the control sample. There is, therefore, no indication of significant prompt quenching of star formation in our sample, in a similar vein to the conclusions of recent work \citep[e.g.][]{Bichanga2024}. Note that this result does not preclude the existence of negative AGN feedback, as this process does not necessarily result in lower instantaneous SFRs during individual star formation episodes \citep[e.g.][]{Ward2022}. Indeed, indicators of quenching tend to correlate more strongly with BH mass, which is a measure of the cumulative output of the AGN over its lifetime \citep[e.g.][]{Piotrowska2022}. 

\subsection{Locations in the cosmic web} 
\label{sec:cosmicweb}

\begin{figure*}
\center
\includegraphics[width=\columnwidth]{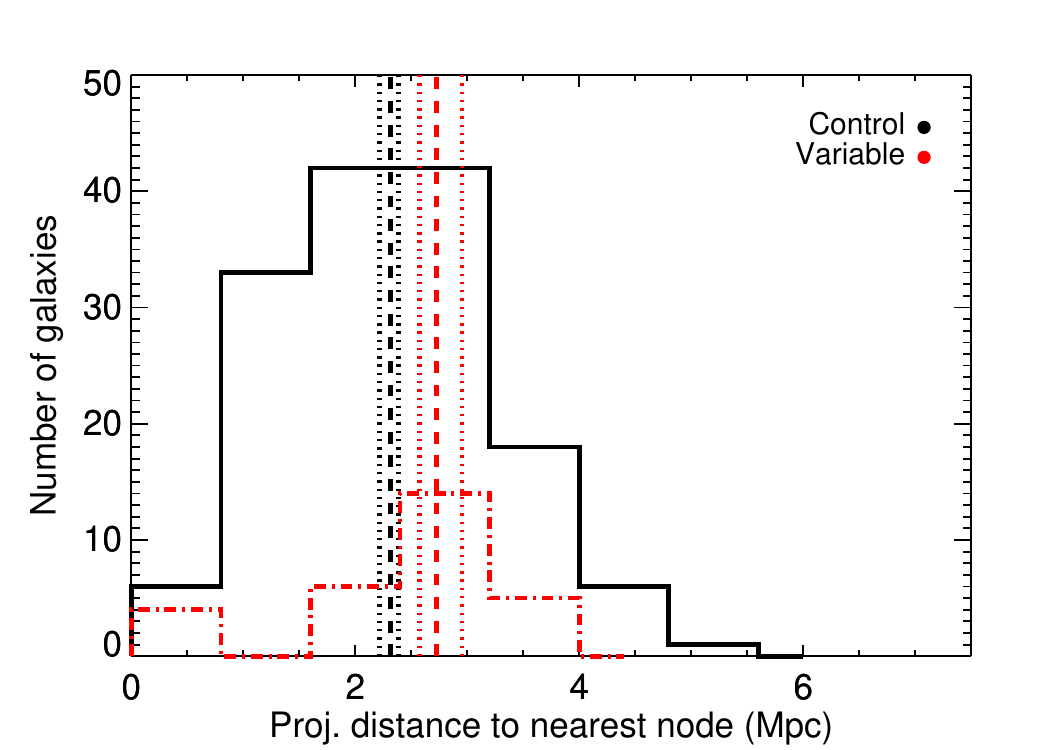}
\includegraphics[width=\columnwidth]{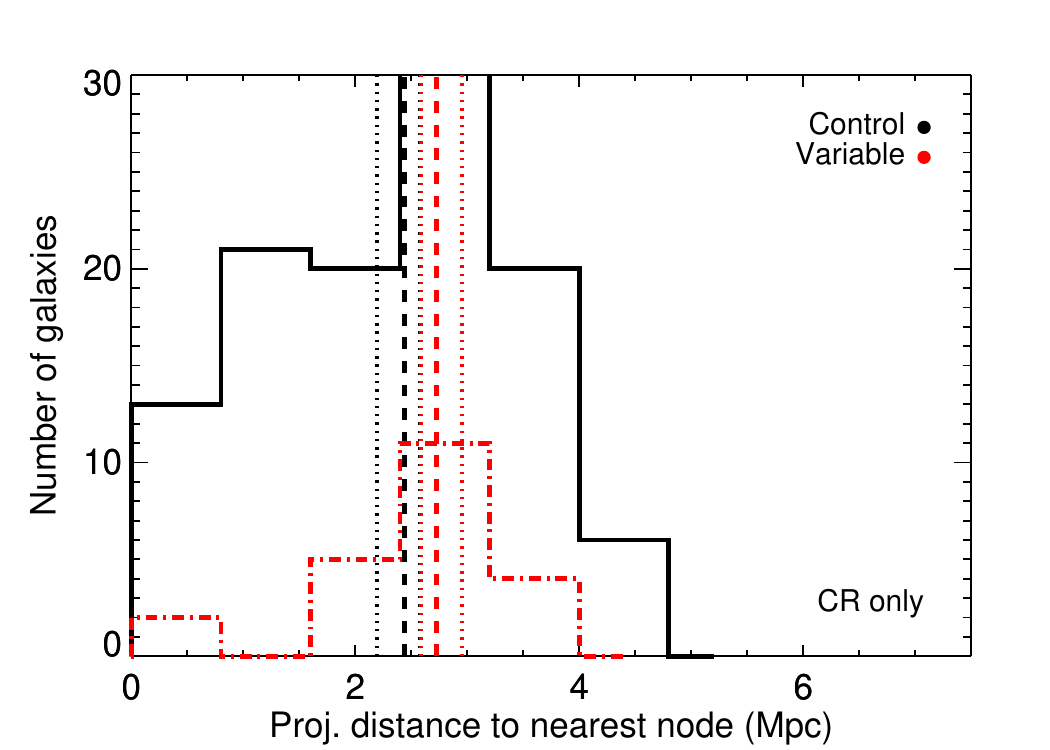}\\
\includegraphics[width=\columnwidth]{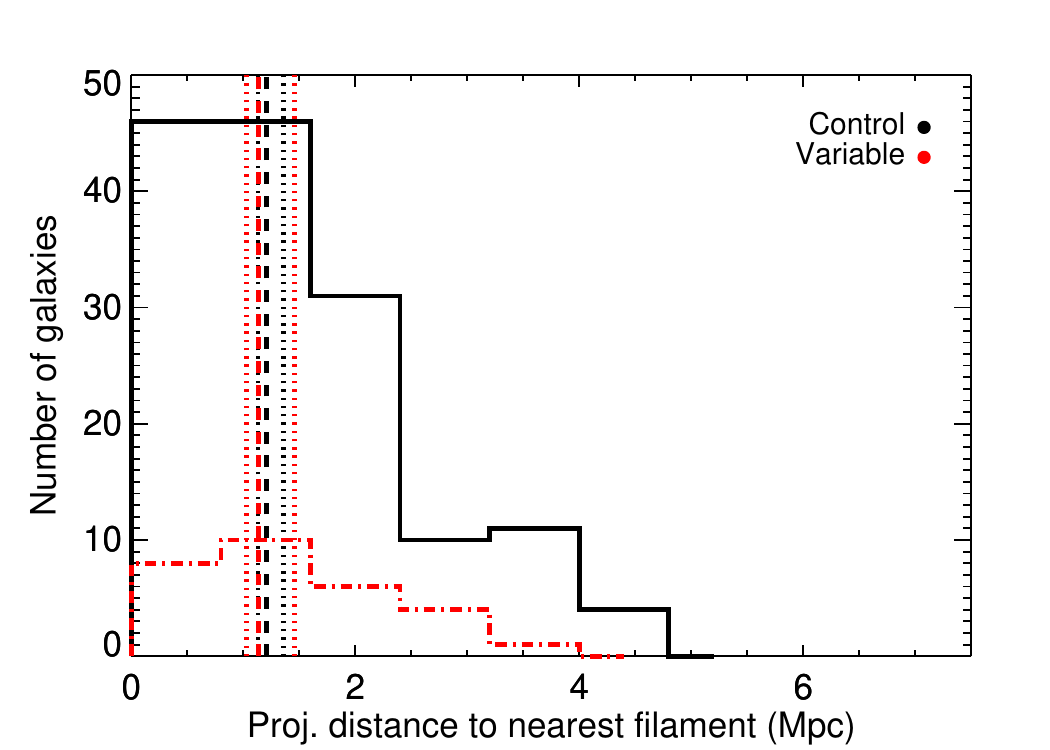}
\includegraphics[width=\columnwidth]{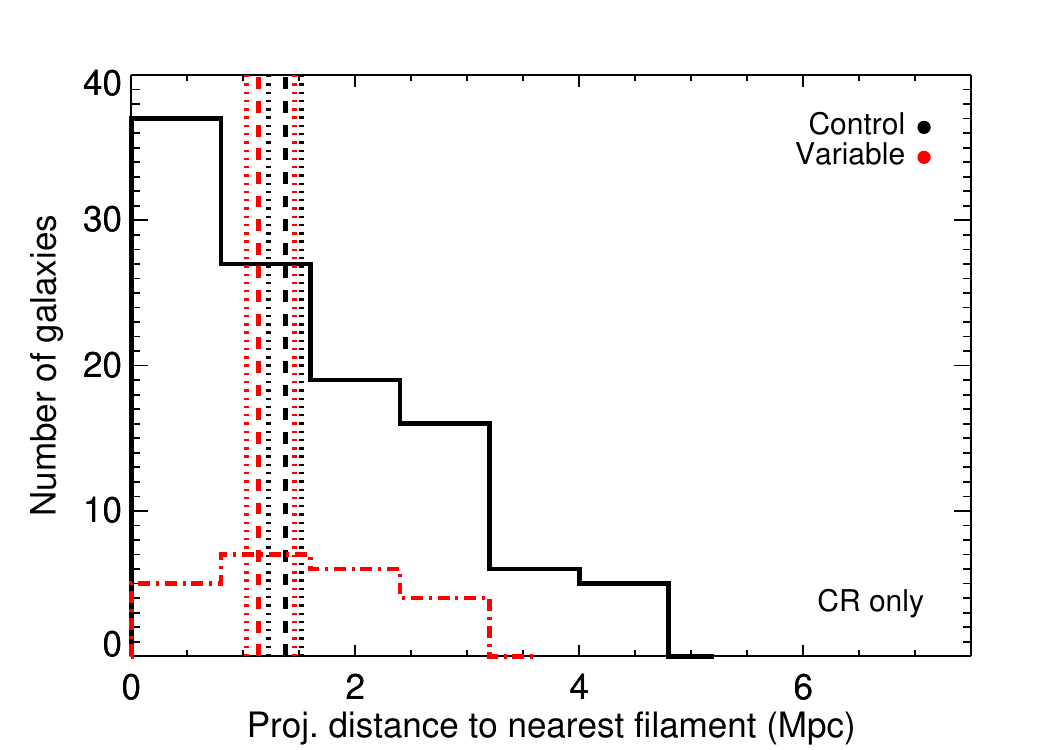}\\
\includegraphics[width=\columnwidth]{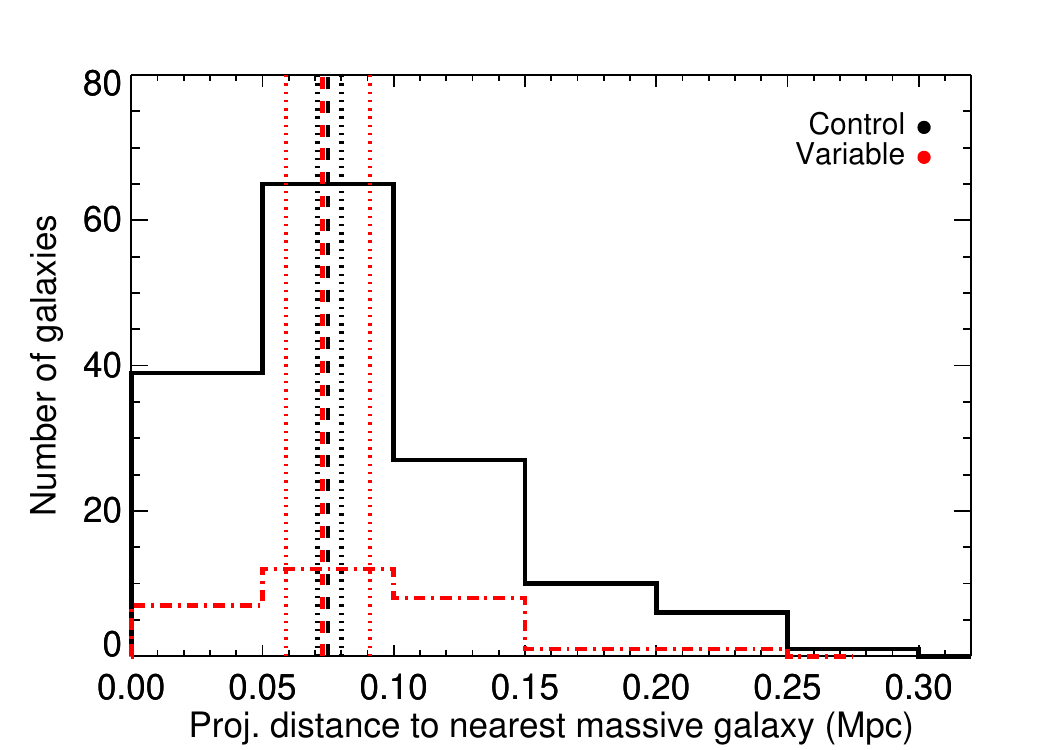}
\includegraphics[width=\columnwidth]{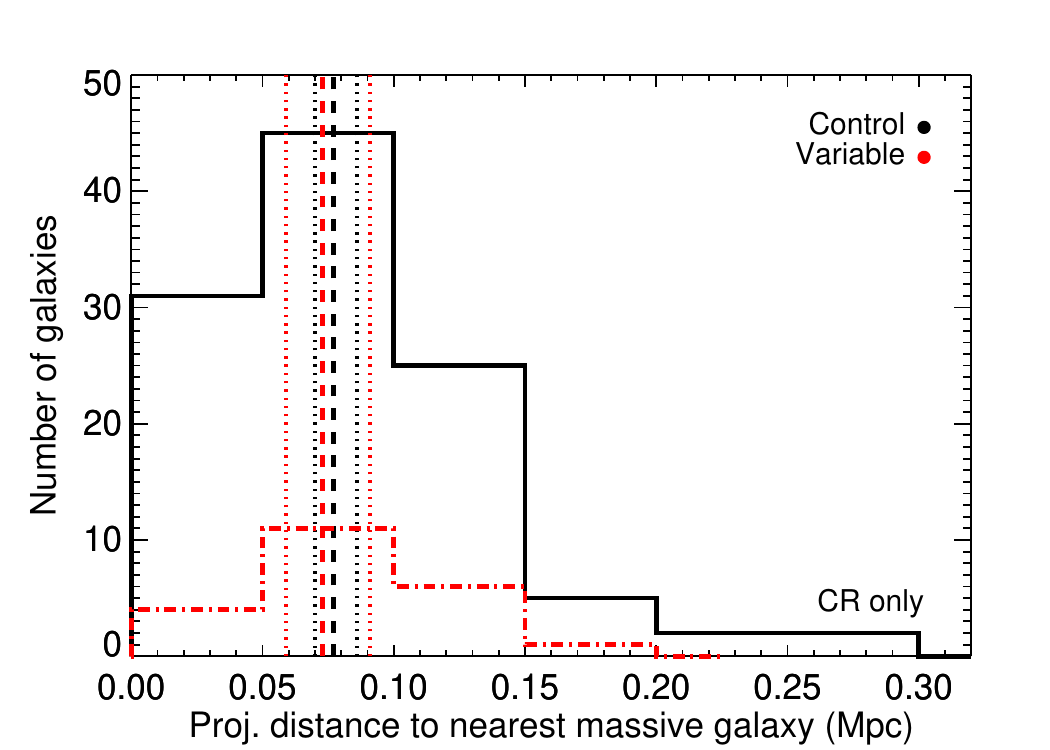}
\caption{Projected distance to the nearest node (top panels), nearest filament (middle panels) and nearest massive galaxy (bottom panels) of dwarf AGN (red) and their control counterparts (black). The left-hand column presents the case where all dwarf AGN are considered, while the distributions in the right-hand column are restricted to galaxies in the VST-COSMOS completeness region (indicated by `CR only' in the panels). Medians and their uncertainties, calculated using bootstrapping, are shown using the dashed and dotted lines respectively.}  
\label{fig:distances}
\end{figure*}

We complete our comparison of the dwarf AGN and control samples by exploring their projected distances to the nearest nodes, filaments and massive galaxies. Figure \ref{fig:distances} shows the distributions of the distances to the nearest nodes (top row), filaments (middle row) and massive galaxies (bottom row) for the AGN (black) and control (red) populations. The left-hand column presents the case where all dwarf AGN are considered, while the distributions in the right-hand column are restricted to galaxies in the VST-COSMOS completeness region (indicated by `CR only' in the panels). Kolmogorov-Smirnov tests for the former case produce p-values of 0.06, 0.92 and 0.89 for the comparison of distances to the nearest nodes (top row), filaments (middle row) and massive galaxies (bottom row) respectively, while the latter case produces p-values of 0.06, 0.73 and 0.83. Recall that p-values less than 0.05 reject the null hypothesis that the distributions are drawn from the same parent distribution. 

While the median values suggest that dwarf AGN may reside at slightly lower distances to nodes than their control counterparts, the KS tests indicate that the dwarf AGN generally appear to reside in similar environments to the control sample. This appears consistent with the findings of other recent studies \citep[e.g.][]{Kristensen2020,Siudek2023,Bichanga2024}, which have employed varying definitions of local environment but also do not find strong differences between the environments of dwarf AGN and the general dwarf population. 

\section{The relative incidence of AGN in the dwarf and massive-galaxy regimes}
\label{sec:frequency}

We complete our study by calculating the relative incidence of AGN in the dwarf and massive-galaxy regimes. As noted above, both host and AGN detectability can impair our ability to calculate the true AGN fraction, particularly in the dwarf regime where both the host galaxy and the AGN can be relatively faint. Therefore, instead of attempting to derive this quantity, we explore the prevalence of AGN in the dwarf regime by estimating the \textit{relative} frequency of AGN in dwarfs and massive galaxies. This allows us to gain insights into the AGN population in dwarfs, given what we already know about the incidence of AGN in the massive-galaxy regime. 

If AGN in dwarfs and massive galaxies are selected using an identical technique, then the ratio of the AGN fractions in the two populations can be used to quantify the relative frequency of AGN in these two mass regimes. Extrapolating this further, since it is thought that the BH occupation fraction of massive galaxies is high, and possibly equal to 1, this ratio may then also provide insights into the BH occupation fraction in dwarf galaxies. 

Given the completeness issues discussed above, we cannot simply calculate the observed AGN fractions in the dwarf and massive-galaxy regimes and divide one by the other. To calculate this ratio we have to take into account, in both mass regimes, (1) the true number of galaxies, (2) how many of those galaxies are actually classified as AGN via variability and (3) the detectability of AGN, given the depth of the VST-COSMOS images and the methodology being employed (in our case, variability). 

In a given stellar mass and redshift range, if $N_{\rm 0}$ is the true number of galaxies, $D$ is the fraction of AGN that are detectable (which we shall use as our definition of detectability) and $N_{\rm  agn}$ is the actual number of AGN detected, then the true AGN fraction ($f_{\rm agn}$) is given by

\begin{equation}
f_{\rm agn} = \frac{N_{\rm agn}}{D.N_{\rm 0}}.
\label{eq:1}
\end{equation}

A trivial consequence of this equation is that the measured number of AGN ($N_{\rm agn}$) leads to an accurate calculation of the true AGN fraction ($f_{\rm agn}$) only if the detectability ($D$) is equal to 1 and the true number of galaxies ($N_{\rm 0}$) is known. 

The detectability ($D$) has two components. The first, which we shall refer to as `host detectability' ($D_{\rm host}$), is driven by the depth of the (typically optical) survey in which the AGN host galaxies are being identified. $D_{\rm host}$ determines which AGN hosts appear in this survey in the first place and, as we show below, can be estimated relatively easily. The second, which we shall refer to as `AGN detectability' ($D_{\rm agn}$), determines the likelihood of an AGN being identified as such by the method being employed. $D_{\rm agn}$ can be more challenging to estimate because it is dependent on the details of the method, how it is being employed and the types of objects being studied. We illustrate this point further using two examples from the recent literature. 

Some AGN-identification methods, such as the commonly-used `BPT' technique that employs optical emission-line ratios \citep[e.g.][]{Baldwin1981,Kewley2001}, are designed to separate star-forming systems from AGN. However, as noted in the introduction (and apparent from Figure \ref{fig:complete}), dwarfs in shallow surveys like the SDSS are biased towards systems which are highly star-forming, particularly outside the very local Universe. Applying this technique to such dwarfs using, for instance, SDSS single-fibre spectroscopy which samples a significant fraction of light around the galaxy centre, leads to AGN fractions close to zero \citep[e.g.][]{Reines2013}. However, if the same technique is applied to SDSS dwarfs in a spatially-resolved manner, via integral field spectroscopy, the measured AGN fractions are significantly higher. \citet{Mezcua2024} have recently explored spatially-resolved emission-line diagnostics, using the [N II]-BPT, [S II]-BPT, [OI]-BPT diagrams \citep{Kewley2001,Kewley2006,Kauffmann2003} augmented by a WHAN analysis \citep{Cidfernandes2010} in the nearby SDSS dwarf population. A conservative classification, which only identifies objects as AGN if they occupy the AGN regions in all three BPT diagrams yields an AGN fraction of $\sim$20 per cent. However, if objects that are classified as star-forming in the [N II]-BPT diagram but AGN in the [S II]-BPT and [OI]-BPT diagrams are also considered to be AGN, then the corresponding AGN fraction rises to $\sim$54 per cent. 

In short, a spatially-resolved analysis of optical emission-line ratios in the dwarf regime suggests (significantly) higher AGN fractions than those derived using single fibre spectroscopy. This discrepancy is likely driven by several reasons. First, while star formation will dilute the contribution of the AGN to the emission lines in both single-fibre and spatially-resolved measurements, the latter is able to isolate the region around the BH where the gas ionisation is dominated by the AGN, increasing the chances of detecting its presence. Second, in many dwarfs the central BHs may wander beyond the 3 arcsec extent of the central SDSS fibre, making it challenging to identify the presence of the AGN. Finally, the accretion disc gets significantly hotter in the dwarf regime, which is likely to cause traditional BPT line ratios to fall outside the parameter space that is typically attributed to AGN \citep{Cann2019}, resulting in an underestimate of the AGN fraction. The basic point here is that the same technique employed in different ways or in different mass regimes can result in very different values of $D_{\rm agn}$. 

In other cases, the ancillary data that is used to identify AGN may restrict the part of the dwarf AGN population that is actually detectable. For example, \citet{Davis2022} have combined deep optical and radio data, from the Hyper Suprime-Cam (HSC) and the Low-Frequency Array (LOFAR) respectively, to study radio AGN in nearby dwarfs. While the depth of the optical HSC data results in a relatively unbiased sample of host galaxies at the redshifts in question, the LOFAR detection limit is close to the peak luminosities of the dwarf radio sources that are LOFAR-detectable. As these peak luminosities are expected to decay to values fainter than the LOFAR detection limit over timescales of a few tens of Myr, the dwarf AGN found by this study are preferentially young. In other words $D_{\rm agn}$ in this sample of dwarfs is a strong function of the age of the AGN. Finding the more numerous population of longer-lived dwarf AGN, which have fainter radio fluxes, requires a deeper radio survey. This will be difficult for the forseeable future, given that the capabilities of the LOFAR surveys will remain unmatched well into the SKA era \citep[e.g.][]{Shimwell2017}. 

In summary, the value of $D_{\rm agn}$ is both difficult to measure and, in many cases, very low, because the properties of the dwarf host galaxies and the details of the AGN identification technique (or both) make it challenging to identify these systems. The final detectability ($D$) in Equation 1 is a combination of $D_{\rm host}$ and $D_{\rm agn}$, i.e. $D = D_{\rm host} \times D_{\rm agn}$. Given the arguments above, it is reasonable to suggest that the extremely low dwarf AGN fractions reported in some studies in the recent literature are produced by low values of detectability, both in terms of $D_{\rm host}$ and $D_{\rm agn}$, and not because AGN do not exist in appreciable numbers in dwarf galaxies. 

If we now wish to compare the AGN fractions in dwarfs and massive galaxies, it follows from Equation 1 that,   

\begin{equation}
\bigg(\frac{f_{\rm agn,d}}{f_{\rm agn,m}}\bigg)= \bigg(\frac{N_{\rm agn,d}}{N_{\rm agn,m}}\bigg) \times \bigg(\frac{N_{\rm 0,m}}{N_{\rm 0,d}}\bigg) \times \bigg(\frac{D_{\rm host,m}}{D_{\rm host,d}}\bigg) \times \bigg(\frac{D_{\rm agn,m}}{D_{\rm agn,d}}\bigg), 
\label{eq:2}
\end{equation}

where `d' and `m' denote dwarfs and massive galaxies respectively. Recall that, in our study, dwarfs are defined as galaxies in the stellar mass range 10$^{8}$ M$_{\odot}$ $<$ M$_{\rm{\star}}$ $<$ 10$^{10}$ M$_{\odot}$, massive galaxies as those in the stellar mass range M$_{\rm{\star}}$ $>$ 10$^{10}$ M$_{\odot}$\footnote{\color{black}The properties of variability-selected AGN in massive galaxies will be presented in a forthcoming paper (Bichang'a et al., in preparation).} and the redshift range of interest is $z<0.4$. 

$N_{\rm agn,d}$ and $N_{\rm agn,m}$ are simply the measured number of AGN in the dwarf and massive-galaxy regimes. $N_{\rm 0,d}$ and $N_{\rm 0,m}$ can be calculated directly from the COSMOS2020 catalogue, since this dataset is complete down to a stellar mass of 10$^{8}$ M$_{\odot}$ and out to $z=0.4$ (Figure \ref{fig:complete}). We therefore take $N_{\rm 0,d}$ and $N_{\rm 0,m}$ to be the numbers of galaxies in the dwarf and massive regimes in COSMOS2020, according to the mass range definitions above. $D_{\rm host,d}$ and $D_{\rm host,m}$ can be derived by calculating the fraction of galaxies, in the dwarf and massive regimes respectively, that are brighter than the magnitude limit of our study ($r=23.5$). Note that, since we are considering fractions in Equation 2, the footprint of the region being considered is not relevant to the calculation. However, cosmic variance may affect the value of $N_{\rm 0,d}$/$N_{\rm 0,m}$, since the shape of the galaxy mass function at low redshift may show some variation with environment \citep[e.g.][]{Papovich2018}. Inserting the relevant values into Equation 2 then yields 

\begin{equation}
\bigg(\frac{f_{\rm agn,d}}{f_{\rm agn,m}}\bigg) = 0.65\:^{+0.52}_{-0.26} \times \bigg(\frac{D_{\rm agn,m}}{D_{\rm agn,d}}\bigg).
\label{eq:3}
\end{equation}



The errors bars are derived from the minimum and maximum values that $f_{\rm agn}$ can have, given the uncertainties on the individual parameters. {\color{black}Note that the error bars on the central value (0.65) are relatively large, which is driven primarily by the sample size}. Finally, we consider the value of $D_{\rm agn,m}$/$D_{\rm agn,d}$. In our case, \citet{Decicco2019} have shown that the fraction of known AGN that are detected via variability increases significantly as the magnitudes of the host galaxies become progressively brighter. This indicates that $D_{\rm agn}$ is larger for more massive galaxies, since they are brighter and therefore ($D_{\rm agn,m}$/$D_{\rm agn,d}$) is greater than 1. Thus, 

\begin{equation}
\bigg(\frac{f_{\rm agn,d}}{f_{\rm agn,m}}\bigg) > 0.65\:^{+0.52}_{-0.26}. 
\label{eq:4}
\end{equation}

{\color{black}This, together with the fact that the central value (0.65) indicates that AGN frequencies are within a factor of 2 of each other, suggests that the incidence of AGN in dwarf and massive galaxies (as defined by the stellar mass ranges above) are likely to be similar.} It is worth noting that if the lower limit of the dwarf stellar mass range is changed from 10$^{8}$ M$_{\odot}$ to 10$^{9}$ M$_{\odot}$, the lower limit on the right hand side in Equation 4 above changes to $0.78\:^{+0.70}_{-0.35}$. The increase in the value of $f_{\rm agn,d}$/$f_{\rm agn,m}$ as we consider more massive dwarfs suggests that the AGN fraction in the dwarf regime may increase as a function of stellar mass, similar to what is thought to happen in the massive-galaxy regime \citep[e.g.][]{Aird2012}. Given the relatively small number of galaxies in our sample, it is difficult to explore this point in more detail because using narrower stellar mass ranges reduces the number of galaxies, making the calculations noisier. It is worth noting here that a similar lack of evolution in the AGN fraction between dwarf and massive galaxies, when observational biases are properly accounted for, has been found via X-ray studies \citep[e.g.][]{Birchall2022}.  


Finally, as mentioned above, the relative incidence of AGN derived above may offer insights into the BH occupation fractions in the two mass regimes. If the BH occupation fraction in massive galaxies is close to unity and the incidence of AGN in dwarfs and massive galaxies is similar then, given the arguments above, it seems reasonable to suggest that the black-hole occupation fraction in dwarf galaxies may also be similarly high. This precludes a scenario in which only the dwarfs which show strong variability actually have BHs, which seems unlikely. 

\section{Summary}
\label{sec:summary}

We have studied 30 nearby ($z<0.4$) dwarf (10$^{8}$ M$_{\odot}$ < M$_{\rm{\star}}$ < 10$^{10}$ M$_{\odot}$) galaxies, which show strong broadband variability over an eleven-year baseline, in the VST-COSMOS survey. We have combined physical parameters from the COSMOS2020 catalogue with visual morphological classifications derived from HST images, to study key properties of our dwarf AGN: AGN flux fractions, morphologies and the role of interactions, SFRs and local environment. We have then compared our dwarf AGN to the general dwarf population using control samples that are matched in stellar mass and redshift. Finally, we have quantified the relative frequencies of AGN in the dwarf and massive regimes in the nearby Universe. Our main conclusions are as follows:

\begin{itemize}

    \item {\color{black}The $r$-band flux fractions contributed by the AGN in our sample are modest (between 5 and 15 per cent of the total flux of the system)}.  
    
    \item The flux fractions in dwarf AGN that are interacting are not larger than in their non-interacting counterparts, suggesting that interactions do not drive higher accretion rates in our dwarf AGN.     

    
     \item The fraction of ETGs is elevated in dwarf AGN compared to the control sample. However, the fraction of interacting galaxies in the AGN and control populations are indistinguishable, indicating that interactions do not play an important role in AGN triggering in our sample. 

    \item Interacting systems only appear within dwarf AGN which are morphologically early-type. This suggests that the presence of interactions makes it more likely that AGN are triggered in dwarf ETGs, while they do not similarly impact the likelihood of AGN triggering in LTGs.

    \item We do not find evidence that AGN activity in our dwarf sample gives rise to significant prompt quenching of star formation activity. 

    \item Dwarf AGN reside at similar distances to nodes, filaments and massive galaxies as their control counterparts, suggesting that AGN triggering in the dwarf regime is not strongly correlated with local environment.  

    \item By combining the true number of galaxies, the detectability of AGN and the measured numbers of variability-selected AGN in dwarf and massive galaxies, we have studied the relative frequency of AGN in these two mass regimes. The incidence of AGN in dwarfs and massive galaxies is likely to be similar (within less than a factor of 2 of each other), with some evidence that the AGN fraction increases with stellar mass in the dwarf regime. 
    
\end{itemize}

Combined with the recent literature, our study suggests that, not only do AGN exist in dwarfs, the incidence of AGN in this regime may be similar to that in massive galaxies. While this empirical result is interesting in its own right, it has strong implications for theoretical models. Given the conclusions of this study (and the recent observational results spanning a range of redshifts), it is unclear whether the lack of BH growth in current cosmological simulations is driven primarily by the characteristics of these simulations, such as their relatively low spatial and mass resolutions, or whether they do in fact faithfully reproduce the behaviour of real dwarf galaxies. It is possible that the implementation of BHs and their interaction with their host galaxies in cosmological simulations will require some revision, informed in part by high-resolution simulations that can probe this process in more detail \citep[e.g.][]{Koudmani2022}. The advent of surveys like the LSST, which will offer optical images of similar depth to that used here, but with a footprint that is four orders of magnitude larger, are keenly anticipated. {\color{black}Studies using such surveys will allow us to study large statistical samples of variability-selected AGN and compare them to more detailed control samples. This is likely to put the results presented here on a firmer statistical footing and offer revolutionary insights into our emerging understanding of the role of AGN in the evolution of dwarf galaxies.} 


\section*{Acknowledgements}

We thank the anonymous referee for many constructive comments that helped us improve the original manuscript. We thank Niel Brandt for many interesting discussions. S Kaviraj, IL and AEW acknowledge support from the STFC (grant numbers ST/Y001257/1 and ST/X001318/1). S. Kaviraj acknowledges a Senior Research Fellowship from Worcester College Oxford. DD acknowledges PON R\&I 2021, CUP E65F21002880003, Fondi di Ricerca di Ateneo (FRA), linea C, progetto TORNADO, and the financial contribution from PRIN-MIUR 2022 and from the Timedomes grant within the ``INAF 2023 Finanziamento della Ricerca Fondamentale''. BB acknowledges a PhD studentship funded by the Centre for Astrophysics Research at the University of Hertfordshire. S. Koudmani is supported by a Research Fellowship from the Royal Commission for the Exhibition of 1851 and a Junior Research Fellowship from St Catharine's College Cambridge. 

The Hyper Suprime-Cam (HSC) collaboration includes the astronomical communities of Japan and Taiwan, and Princeton University. The HSC instrumentation and software were developed by the National Astronomical Observatory of Japan (NAOJ), the Kavli Institute for the Physics and Mathematics of the Universe (Kavli IPMU), the University of Tokyo, the High Energy Accelerator Research Organization (KEK), the Academia Sinica Institute for Astronomy and Astrophysics in Taiwan (ASIAA), and Princeton University. Funding was contributed by the FIRST program from the Japanese Cabinet Office, the Ministry of Education, Culture, Sports, Science and Technology (MEXT), the Japan Society for the Promotion of Science (JSPS), Japan Science and Technology Agency (JST), the Toray Science Foundation, NAOJ, Kavli IPMU, KEK, ASIAA, and Princeton University. 

This paper makes use of software developed for Vera C. Rubin Observatory. We thank the Rubin Observatory for making their code available as free software at http://pipelines.lsst.io/. This paper is based on data collected at the Subaru Telescope and retrieved from the HSC data archive system, which is operated by the Subaru Telescope and Astronomy Data Center (ADC) at NAOJ. Data analysis was in part carried out with the cooperation of Center for Computational Astrophysics (CfCA), NAOJ. We are honored and grateful for the opportunity of observing the Universe from Maunakea, which has the cultural, historical and natural significance in Hawaii. 


\section*{Data Availability}
The galaxy sample used in this study has been constructed using the catalogue from \citet{Weaver2022}. The density maps were created using the DisPerSE algorithm which is described in \citet{Sousbie2011}.


\bibliographystyle{mnras}
\bibliography{references} 

\begin{thebibliography}{}
\makeatletter
\relax
\def\mn@urlcharsother{\let\do\@makeother \do\$\do\&\do\#\do\^\do\_\do\%\do\~}
\def\mn@doi{\begingroup\mn@urlcharsother \@ifnextchar [ {\mn@doi@} {\mn@doi@[]}}
\def\mn@doi@[#1]#2{\def\@tempa{#1}\ifx\@tempa\@empty \href {http://dx.doi.org/#2} {doi:#2}\else \href {http://dx.doi.org/#2} {#1}\fi \endgroup}
\def\mn@eprint#1#2{\mn@eprint@#1:#2::\@nil}
\def\mn@eprint@arXiv#1{\href {http://arxiv.org/abs/#1} {{\tt arXiv:#1}}}
\def\mn@eprint@dblp#1{\href {http://dblp.uni-trier.de/rec/bibtex/#1.xml} {dblp:#1}}
\def\mn@eprint@#1:#2:#3:#4\@nil{\def\@tempa {#1}\def\@tempb {#2}\def\@tempc {#3}\ifx \@tempc \@empty \let \@tempc \@tempb \let \@tempb \@tempa \fi \ifx \@tempb \@empty \def\@tempb {arXiv}\fi \@ifundefined {mn@eprint@\@tempb}{\@tempb:\@tempc}{\expandafter \expandafter \csname mn@eprint@\@tempb\endcsname \expandafter{\@tempc}}}

\bibitem[\protect\citeauthoryear{{Aihara} et~al.,}{{Aihara} et~al.}{2019}]{Aihara2019}
{Aihara} H.,  et~al., 2019, \mn@doi [\pasj] {10.1093/pasj/psz103}, \href {https://ui.adsabs.harvard.edu/abs/2019PASJ...71..114A} {71, 114}

\bibitem[\protect\citeauthoryear{{Aird} et~al.,}{{Aird} et~al.}{2012}]{Aird2012}
{Aird} J.,  et~al., 2012, \mn@doi [\apj] {10.1088/0004-637X/746/1/90}, \href {https://ui.adsabs.harvard.edu/abs/2012ApJ...746...90A} {746, 90}

\bibitem[\protect\citeauthoryear{{Alam} et~al.,}{{Alam} et~al.}{2015}]{Alam2015}
{Alam} S.,  et~al., 2015, \mn@doi [\apjs] {10.1088/0067-0049/219/1/12}, \href {https://ui.adsabs.harvard.edu/abs/2015ApJS..219...12A} {219, 12}

\bibitem[\protect\citeauthoryear{{Angl{\'e}s-Alc{\'a}zar}, {Faucher-Gigu{\`e}re}, {Quataert}, {Hopkins}, {Feldmann}, {Torrey}, {Wetzel}  \& {Kere{\v{s}}}}{{Angl{\'e}s-Alc{\'a}zar} et~al.}{2017}]{Angles-Alcazar2017}
{Angl{\'e}s-Alc{\'a}zar} D.,  {Faucher-Gigu{\`e}re} C.-A.,  {Quataert} E.,  {Hopkins} P.~F.,  {Feldmann} R.,  {Torrey} P.,  {Wetzel} A.,   {Kere{\v{s}}} D.,  2017, \mn@doi [\mnras] {10.1093/mnrasl/slx161}, \href {https://ui.adsabs.harvard.edu/abs/2017MNRAS.472L.109A} {472, L109}

\bibitem[\protect\citeauthoryear{{Arjona-Galvez}, {Di Cintio}  \& {Grand}}{{Arjona-Galvez} et~al.}{2024}]{ArjonaGalvez2024}
{Arjona-Galvez} E.,  {Di Cintio} A.,   {Grand} R. J.~J.,  2024, \mn@doi [arXiv e-prints] {10.48550/arXiv.2402.00929}, \href {https://ui.adsabs.harvard.edu/abs/2024arXiv240200929A} {p. arXiv:2402.00929}

\bibitem[\protect\citeauthoryear{{Arnouts} et~al.,}{{Arnouts} et~al.}{2002}]{Arnouts2002}
{Arnouts} S.,  et~al., 2002, \mn@doi [\mnras] {10.1046/j.1365-8711.2002.04988.x}, \href {https://ui.adsabs.harvard.edu/abs/2002MNRAS.329..355A} {329, 355}

\bibitem[\protect\citeauthoryear{{Ashby} et~al.,}{{Ashby} et~al.}{2013}]{Ashby2013}
{Ashby} M.~L.~N.,  et~al., 2013, \mn@doi [\apj] {10.1088/0004-637X/769/1/80}, \href {https://ui.adsabs.harvard.edu/abs/2013ApJ...769...80A} {769, 80}

\bibitem[\protect\citeauthoryear{{Ashby} et~al.,}{{Ashby} et~al.}{2015}]{Ashby2015}
{Ashby} M.~L.~N.,  et~al., 2015, \mn@doi [\apjs] {10.1088/0067-0049/218/2/33}, \href {https://ui.adsabs.harvard.edu/abs/2015ApJS..218...33A} {218, 33}

\bibitem[\protect\citeauthoryear{{Ashby} et~al.,}{{Ashby} et~al.}{2018}]{Ashby2018}
{Ashby} M.~L.~N.,  et~al., 2018, \mn@doi [\apjs] {10.3847/1538-4365/aad4fb}, \href {https://ui.adsabs.harvard.edu/abs/2018ApJS..237...39A} {237, 39}

\bibitem[\protect\citeauthoryear{{Baker} et~al.,}{{Baker} et~al.}{2025}]{Baker2025}
{Baker} W.~M.,  et~al., 2025, \mn@doi [\aap] {10.1051/0004-6361/202553766}, \href {https://ui.adsabs.harvard.edu/abs/2025A&A...697A..90B} {697, A90}

\bibitem[\protect\citeauthoryear{{Baldassare}, {Geha}  \& {Greene}}{{Baldassare} et~al.}{2020a}]{Baldassare2020b}
{Baldassare} V.~F.,  {Geha} M.,   {Greene} J.,  2020a, \mn@doi [\apj] {10.3847/1538-4357/ab8936}, \href {https://ui.adsabs.harvard.edu/abs/2020ApJ...896...10B} {896, 10}

\bibitem[\protect\citeauthoryear{{Baldassare}, {Dickey}, {Geha}  \& {Reines}}{{Baldassare} et~al.}{2020b}]{Baldassare2020a}
{Baldassare} V.~F.,  {Dickey} C.,  {Geha} M.,   {Reines} A.~E.,  2020b, \mn@doi [\apjl] {10.3847/2041-8213/aba0c1}, \href {https://ui.adsabs.harvard.edu/abs/2020ApJ...898L...3B} {898, L3}

\bibitem[\protect\citeauthoryear{{Baldwin}, {Phillips}  \& {Terlevich}}{{Baldwin} et~al.}{1981}]{Baldwin1981}
{Baldwin} J.~A.,  {Phillips} M.~M.,   {Terlevich} R.,  1981, \mn@doi [\pasp] {10.1086/130766}, \href {https://ui.adsabs.harvard.edu/abs/1981PASP...93....5B} {93, 5}

\bibitem[\protect\citeauthoryear{{Barai} \& {de Gouveia Dal Pino}}{{Barai} \& {de Gouveia Dal Pino}}{2019}]{Barai2019}
{Barai} P.,  {de Gouveia Dal Pino} E.~M.,  2019, \mn@doi [\mnras] {10.1093/mnras/stz1616}, \href {https://ui.adsabs.harvard.edu/abs/2019MNRAS.487.5549B} {487, 5549}

\bibitem[\protect\citeauthoryear{{Beckmann} et~al.,}{{Beckmann} et~al.}{2017}]{Beckmann2017}
{Beckmann} R.~S.,  et~al., 2017, \mn@doi [\mnras] {10.1093/mnras/stx1831}, \href {http://adsabs.harvard.edu/abs/2017MNRAS.472..949B} {472, 949}

\bibitem[\protect\citeauthoryear{{Behroozi}, {Wechsler}  \& {Conroy}}{{Behroozi} et~al.}{2013}]{Behroozi2013}
{Behroozi} P.~S.,  {Wechsler} R.~H.,   {Conroy} C.,  2013, \mn@doi [\apj] {10.1088/0004-637X/770/1/57}, \href {https://ui.adsabs.harvard.edu/abs/2013ApJ...770...57B} {770, 57}

\bibitem[\protect\citeauthoryear{{Bellovary} et~al.,}{{Bellovary} et~al.}{2021}]{Bellovary2021}
{Bellovary} J.~M.,  et~al., 2021, \mn@doi [\mnras] {10.1093/mnras/stab1665}, \href {https://ui.adsabs.harvard.edu/abs/2021MNRAS.505.5129B} {505, 5129}

\bibitem[\protect\citeauthoryear{{Bichang'a}, {Kaviraj}, {Lazar}, {Jackson}, {Das}, {Smith}, {Watkins}  \& {Martin}}{{Bichang'a} et~al.}{2024}]{Bichanga2024}
{Bichang'a} B.,  {Kaviraj} S.,  {Lazar} I.,  {Jackson} R.~A.,  {Das} S.,  {Smith} D.~J.~B.,  {Watkins} A.~E.,   {Martin} G.,  2024, \mn@doi [\mnras] {10.1093/mnras/stae1441}, \href {https://ui.adsabs.harvard.edu/abs/2024MNRAS.532..613B} {532, 613}

\bibitem[\protect\citeauthoryear{{Birchall}, {Watson}  \& {Aird}}{{Birchall} et~al.}{2020}]{Birchall2020}
{Birchall} K.~L.,  {Watson} M.~G.,   {Aird} J.,  2020, \mn@doi [\mnras] {10.1093/mnras/staa040}, \href {https://ui.adsabs.harvard.edu/abs/2020MNRAS.492.2268B} {492, 2268}

\bibitem[\protect\citeauthoryear{{Birchall}, {Watson}, {Aird}  \& {Starling}}{{Birchall} et~al.}{2022}]{Birchall2022}
{Birchall} K.~L.,  {Watson} M.~G.,  {Aird} J.,   {Starling} R.~L.~C.,  2022, \mn@doi [\mnras] {10.1093/mnras/stab3573}, \href {https://ui.adsabs.harvard.edu/abs/2022MNRAS.510.4556B} {510, 4556}

\bibitem[\protect\citeauthoryear{{Bruzual} \& {Charlot}}{{Bruzual} \& {Charlot}}{2003}]{Bruzual2003}
{Bruzual} G.,  {Charlot} S.,  2003, \mn@doi [\mnras] {10.1046/j.1365-8711.2003.06897.x}, \href {https://ui.adsabs.harvard.edu/#abs/2003MNRAS.344.1000B} {344, 1000}

\bibitem[\protect\citeauthoryear{{Bundy} et~al.,}{{Bundy} et~al.}{2015}]{Bundy2015}
{Bundy} K.,  et~al., 2015, \mn@doi [\apj] {10.1088/0004-637X/798/1/7}, \href {https://ui.adsabs.harvard.edu/abs/2015ApJ...798....7B} {798, 7}

\bibitem[\protect\citeauthoryear{{Burke} et~al.,}{{Burke} et~al.}{2022}]{Burke2022}
{Burke} C.~J.,  et~al., 2022, \mn@doi [\mnras] {10.1093/mnras/stac2262}, \href {https://ui.adsabs.harvard.edu/abs/2022MNRAS.516.2736B} {516, 2736}

\bibitem[\protect\citeauthoryear{{Burke}, {Liu}, {Ward}, {Liu}, {Natarajan}  \& {Greene}}{{Burke} et~al.}{2024}]{Burke2024}
{Burke} C.~J.,  {Liu} Y.,  {Ward} C.~A.,  {Liu} X.,  {Natarajan} P.,   {Greene} J.~E.,  2024, \mn@doi [\apj] {10.3847/1538-4357/ad54ca}, \href {https://ui.adsabs.harvard.edu/abs/2024ApJ...971..140B} {971, 140}

\bibitem[\protect\citeauthoryear{{Buta}, {Mitra}, {de Vaucouleurs}  \& {Corwin}}{{Buta} et~al.}{1994}]{Buta1994}
{Buta} R.,  {Mitra} S.,  {de Vaucouleurs} G.,   {Corwin} H.~G. J.,  1994, \mn@doi [\aj] {10.1086/116838}, \href {https://ui.adsabs.harvard.edu/abs/1994AJ....107..118B} {107, 118}

\bibitem[\protect\citeauthoryear{{Cameron}}{{Cameron}}{2011}]{Cameron2011}
{Cameron} E.,  2011, \mn@doi [\pasa] {10.1071/AS10046}, \href {https://ui.adsabs.harvard.edu/abs/2011PASA...28..128C} {28, 128}

\bibitem[\protect\citeauthoryear{{Cann}, {Satyapal}, {Abel}, {Blecha}, {Mushotzky}, {Reynolds}  \& {Secrest}}{{Cann} et~al.}{2019}]{Cann2019}
{Cann} J.~M.,  {Satyapal} S.,  {Abel} N.~P.,  {Blecha} L.,  {Mushotzky} R.~F.,  {Reynolds} C.~S.,   {Secrest} N.~J.,  2019, \mn@doi [\apjl] {10.3847/2041-8213/aaf88d}, \href {https://ui.adsabs.harvard.edu/abs/2019ApJ...870L...2C} {870, L2}

\bibitem[\protect\citeauthoryear{{Capaccioli} \& {Schipani}}{{Capaccioli} \& {Schipani}}{2011}]{Capaccioli2011}
{Capaccioli} M.,  {Schipani} P.,  2011, The Messenger, \href {http://adsabs.harvard.edu/abs/2011Msngr.146....2C} {146, 2}

\bibitem[\protect\citeauthoryear{{Chen} et~al.,}{{Chen} et~al.}{2017}]{Chen2017}
{Chen} C. T.~J.,  et~al., 2017, \mn@doi [\apj] {10.3847/1538-4357/aa5d5b}, \href {https://ui.adsabs.harvard.edu/abs/2017ApJ...837...48C} {837, 48}

\bibitem[\protect\citeauthoryear{{Cid Fernandes}, {Stasi{\'n}ska}, {Schlickmann}, {Mateus}, {Vale Asari}, {Schoenell}  \& {Sodr{\'e}}}{{Cid Fernandes} et~al.}{2010}]{Cidfernandes2010}
{Cid Fernandes} R.,  {Stasi{\'n}ska} G.,  {Schlickmann} M.~S.,  {Mateus} A.,  {Vale Asari} N.,  {Schoenell} W.,   {Sodr{\'e}} L.,  2010, \mn@doi [\mnras] {10.1111/j.1365-2966.2009.16185.x}, \href {https://ui.adsabs.harvard.edu/abs/2010MNRAS.403.1036C} {403, 1036}

\bibitem[\protect\citeauthoryear{{Conselice}}{{Conselice}}{2003}]{Conselice2003}
{Conselice} C.~J.,  2003, \mn@doi [\apjs] {10.1086/375001}, \href {https://ui.adsabs.harvard.edu/abs/2003ApJS..147....1C} {147, 1}

\bibitem[\protect\citeauthoryear{{Croton} et~al.,}{{Croton} et~al.}{2006}]{Croton2006}
{Croton} D.~J.,  et~al., 2006, \mn@doi [\mnras] {10.1111/j.1365-2966.2005.09675.x}, \href {http://adsabs.harvard.edu/abs/2006MNRAS.365...11C} {365, 11}

\bibitem[\protect\citeauthoryear{{Davis} et~al.,}{{Davis} et~al.}{2020}]{Davis2020}
{Davis} T.~A.,  et~al., 2020, \mn@doi [\mnras] {10.1093/mnras/staa1567}, \href {https://ui.adsabs.harvard.edu/abs/2020MNRAS.496.4061D} {496, 4061}

\bibitem[\protect\citeauthoryear{{Davis} et~al.,}{{Davis} et~al.}{2022}]{Davis2022}
{Davis} F.,  et~al., 2022, \mn@doi [\mnras] {10.1093/mnras/stac068}, \href {https://ui.adsabs.harvard.edu/abs/2022MNRAS.511.4109D} {511, 4109}

\bibitem[\protect\citeauthoryear{{De Cicco} et~al.,}{{De Cicco} et~al.}{2015}]{Decicco2015}
{De Cicco} D.,  et~al., 2015, \mn@doi [\aap] {10.1051/0004-6361/201424906}, \href {https://ui.adsabs.harvard.edu/abs/2015A&A...574A.112D} {574, A112}

\bibitem[\protect\citeauthoryear{{De Cicco} et~al.,}{{De Cicco} et~al.}{2019}]{Decicco2019}
{De Cicco} D.,  et~al., 2019, \mn@doi [\aap] {10.1051/0004-6361/201935659}, \href {https://ui.adsabs.harvard.edu/abs/2019A&A...627A..33D} {627, A33}

\bibitem[\protect\citeauthoryear{{Dickey}, {Geha}, {Wetzel}  \& {El-Badry}}{{Dickey} et~al.}{2019}]{Dickey2019}
{Dickey} C.~M.,  {Geha} M.,  {Wetzel} A.,   {El-Badry} K.,  2019, \mn@doi [\apj] {10.3847/1538-4357/ab3220}, \href {https://ui.adsabs.harvard.edu/abs/2019ApJ...884..180D} {884, 180}

\bibitem[\protect\citeauthoryear{{Drinkwater}, {Gregg}  \& {Colless}}{{Drinkwater} et~al.}{2001}]{Drinkwater2001}
{Drinkwater} M.~J.,  {Gregg} M.~D.,   {Colless} M.,  2001, \mn@doi [\apjl] {10.1086/319113}, \href {https://ui.adsabs.harvard.edu/abs/2001ApJ...548L.139D} {548, L139}

\bibitem[\protect\citeauthoryear{{Dubois}, {Volonteri}, {Silk}, {Devriendt}, {Slyz}  \& {Teyssier}}{{Dubois} et~al.}{2015}]{Dubois2015}
{Dubois} Y.,  {Volonteri} M.,  {Silk} J.,  {Devriendt} J.,  {Slyz} A.,   {Teyssier} R.,  2015, \mn@doi [\mnras] {10.1093/mnras/stv1416}, \href {https://ui.adsabs.harvard.edu/abs/2015MNRAS.452.1502D} {452, 1502}

\bibitem[\protect\citeauthoryear{{Dubois} et~al.,}{{Dubois} et~al.}{2021}]{Dubois2021}
{Dubois} Y.,  et~al., 2021, \mn@doi [\aap] {10.1051/0004-6361/202039429}, \href {https://ui.adsabs.harvard.edu/abs/2021A&A...651A.109D} {651, A109}

\bibitem[\protect\citeauthoryear{{Duc} et~al.,}{{Duc} et~al.}{2015}]{Duc2015}
{Duc} P.-A.,  et~al., 2015, \mn@doi [\mnras] {10.1093/mnras/stu2019}, \href {https://ui.adsabs.harvard.edu/abs/2015MNRAS.446..120D} {446, 120}

\bibitem[\protect\citeauthoryear{{Elmer}, {Almaini}, {Merrifield}, {Hartley}, {Maltby}, {Lawrence}, {Botti}  \& {Hirst}}{{Elmer} et~al.}{2020}]{Elmer2020}
{Elmer} E.,  {Almaini} O.,  {Merrifield} M.,  {Hartley} W.~G.,  {Maltby} D.~T.,  {Lawrence} A.,  {Botti} I.,   {Hirst} P.,  2020, \mn@doi [\mnras] {10.1093/mnras/staa381}, \href {https://ui.adsabs.harvard.edu/abs/2020MNRAS.493.3026E} {493, 3026}

\bibitem[\protect\citeauthoryear{{Endsley} et~al.,}{{Endsley} et~al.}{2024}]{Endsley2024}
{Endsley} R.,  et~al., 2024, \mn@doi [\mnras] {10.1093/mnras/stae1857}, \href {https://ui.adsabs.harvard.edu/abs/2024MNRAS.533.1111E} {533, 1111}

\bibitem[\protect\citeauthoryear{{Er{\'o}stegui}, {Mezcua}, {Siudek}, {Dom{\'\i}nguez S{\'a}nchez}  \& {Rodr{\'\i}guez Morales}}{{Er{\'o}stegui} et~al.}{2025}]{Erostegui2025}
{Er{\'o}stegui} A.,  {Mezcua} M.,  {Siudek} M.,  {Dom{\'\i}nguez S{\'a}nchez} H.,   {Rodr{\'\i}guez Morales} V.,  2025, \mn@doi [\aap] {10.1051/0004-6361/202554387}, \href {https://ui.adsabs.harvard.edu/abs/2025A&A...699A.330E} {699, A330}

\bibitem[\protect\citeauthoryear{{Fabian}}{{Fabian}}{2012}]{Fabian2012}
{Fabian} A.~C.,  2012, \mn@doi [\araa] {10.1146/annurev-astro-081811-125521}, \href {https://ui.adsabs.harvard.edu/abs/2012ARA&A..50..455F} {50, 455}

\bibitem[\protect\citeauthoryear{{Finoguenov} et~al.,}{{Finoguenov} et~al.}{2007}]{Finoguenov2007}
{Finoguenov} A.,  et~al., 2007, \mn@doi [\apjs] {10.1086/516577}, \href {https://ui.adsabs.harvard.edu/abs/2007ApJS..172..182F} {172, 182}

\bibitem[\protect\citeauthoryear{{Fouqu{\'e}}, {Solanes}, {Sanchis}  \& {Balkowski}}{{Fouqu{\'e}} et~al.}{2001}]{Fouque2001}
{Fouqu{\'e}} P.,  {Solanes} J.~M.,  {Sanchis} T.,   {Balkowski} C.,  2001, \mn@doi [\aap] {10.1051/0004-6361:20010833}, \href {https://ui.adsabs.harvard.edu/abs/2001A&A...375..770F} {375, 770}

\bibitem[\protect\citeauthoryear{{Gardner} et~al.,}{{Gardner} et~al.}{2006}]{Gardner2006}
{Gardner} J.~P.,  et~al., 2006, \mn@doi [\ssr] {10.1007/s11214-006-8315-7}, \href {http://adsabs.harvard.edu/abs/2006SSRv..123..485G} {123, 485}

\bibitem[\protect\citeauthoryear{{Gavazzi}, {Adami}, {Durret}, {Cuillandre}, {Ilbert}, {Mazure}, {Pell{\'o}}  \& {Ulmer}}{{Gavazzi} et~al.}{2009}]{Gavazzi2009}
{Gavazzi} R.,  {Adami} C.,  {Durret} F.,  {Cuillandre} J.~C.,  {Ilbert} O.,  {Mazure} A.,  {Pell{\'o}} R.,   {Ulmer} M.~P.,  2009, \mn@doi [\aap] {10.1051/0004-6361/200911841}, \href {https://ui.adsabs.harvard.edu/abs/2009A&A...498L..33G} {498, L33}

\bibitem[\protect\citeauthoryear{{Geha} et~al.,}{{Geha} et~al.}{2017}]{Geha2017}
{Geha} M.,  et~al., 2017, \mn@doi [\apj] {10.3847/1538-4357/aa8626}, \href {https://ui.adsabs.harvard.edu/abs/2017ApJ...847....4G} {847, 4}

\bibitem[\protect\citeauthoryear{{George} et~al.,}{{George} et~al.}{2011}]{George2011}
{George} M.~R.,  et~al., 2011, \mn@doi [\apj] {10.1088/0004-637X/742/2/125}, \href {https://ui.adsabs.harvard.edu/abs/2011ApJ...742..125G} {742, 125}

\bibitem[\protect\citeauthoryear{{Gozaliasl} et~al.,}{{Gozaliasl} et~al.}{2014}]{Gozaliasl2014}
{Gozaliasl} G.,  et~al., 2014, \mn@doi [\aap] {10.1051/0004-6361/201322459}, \href {https://ui.adsabs.harvard.edu/abs/2014A&A...566A.140G} {566, A140}

\bibitem[\protect\citeauthoryear{{Gozaliasl} et~al.,}{{Gozaliasl} et~al.}{2019}]{Gozaliasl2019}
{Gozaliasl} G.,  et~al., 2019, \mn@doi [\mnras] {10.1093/mnras/sty3203}, \href {https://ui.adsabs.harvard.edu/abs/2019MNRAS.483.3545G} {483, 3545}

\bibitem[\protect\citeauthoryear{{Greene} \& {Ho}}{{Greene} \& {Ho}}{2007}]{Greene2007}
{Greene} J.~E.,  {Ho} L.~C.,  2007, \mn@doi [\apj] {10.1086/522082}, \href {https://ui.adsabs.harvard.edu/abs/2007ApJ...670...92G} {670, 92}

\bibitem[\protect\citeauthoryear{{Habouzit}, {Volonteri}  \& {Dubois}}{{Habouzit} et~al.}{2017}]{Habouzit2017}
{Habouzit} M.,  {Volonteri} M.,   {Dubois} Y.,  2017, \mn@doi [\mnras] {10.1093/mnras/stx666}, \href {https://ui.adsabs.harvard.edu/abs/2017MNRAS.468.3935H} {468, 3935}

\bibitem[\protect\citeauthoryear{{Ilbert} et~al.,}{{Ilbert} et~al.}{2006}]{Ilbert2006}
{Ilbert} O.,  et~al., 2006, \mn@doi [\aap] {10.1051/0004-6361:20065138}, \href {https://ui.adsabs.harvard.edu/abs/2006A&A...457..841I} {457, 841}

\bibitem[\protect\citeauthoryear{{Ivezi{\'c}} et~al.,}{{Ivezi{\'c}} et~al.}{2019}]{Ivezic2019}
{Ivezi{\'c}} {\v{Z}}.,  et~al., 2019, \mn@doi [\apj] {10.3847/1538-4357/ab042c}, \href {https://ui.adsabs.harvard.edu/abs/2019ApJ...873..111I} {873, 111}

\bibitem[\protect\citeauthoryear{{Jarrett} et~al.,}{{Jarrett} et~al.}{2011}]{Jarrett2011}
{Jarrett} T.~H.,  et~al., 2011, \mn@doi [\apj] {10.1088/0004-637X/735/2/112}, \href {http://adsabs.harvard.edu/abs/2011ApJ...735..112J} {735, 112}

\bibitem[\protect\citeauthoryear{{Juod{\v{z}}balis} et~al.,}{{Juod{\v{z}}balis} et~al.}{2023}]{Juodzbalis2023}
{Juod{\v{z}}balis} I.,  et~al., 2023, \mn@doi [\mnras] {10.1093/mnras/stad2396}, \href {https://ui.adsabs.harvard.edu/abs/2023MNRAS.525.1353J} {525, 1353}

\bibitem[\protect\citeauthoryear{{Kauffmann} et~al.,}{{Kauffmann} et~al.}{2003}]{Kauffmann2003}
{Kauffmann} G.,  et~al., 2003, \mn@doi [\mnras] {10.1046/j.1365-8711.2003.06292.x}, \href {https://ui.adsabs.harvard.edu/abs/2003MNRAS.341...54K} {341, 54}

\bibitem[\protect\citeauthoryear{{Kaviraj}}{{Kaviraj}}{2014}]{Kaviraj2014b}
{Kaviraj} S.,  2014, \mn@doi [\mnras] {10.1093/mnras/stu338}, \href {http://adsabs.harvard.edu/abs/2014MNRAS.440.2944K} {440, 2944}

\bibitem[\protect\citeauthoryear{{Kaviraj} et~al.,}{{Kaviraj} et~al.}{2017}]{Kaviraj2017}
{Kaviraj} S.,  et~al., 2017, \mn@doi [\mnras] {10.1093/mnras/stx126}, \href {http://adsabs.harvard.edu/abs/2017MNRAS.467.4739K} {467, 4739}

\bibitem[\protect\citeauthoryear{{Kaviraj}, {Martin}  \& {Silk}}{{Kaviraj} et~al.}{2019}]{Kaviraj2019}
{Kaviraj} S.,  {Martin} G.,   {Silk} J.,  2019, \mn@doi [\mnras] {10.1093/mnrasl/slz102}, \href {https://ui.adsabs.harvard.edu/abs/2019MNRAS.489L..12K} {489, L12}

\bibitem[\protect\citeauthoryear{{Kaviraj}, {Lazar}, {Watkins}, {Laigle}, {Martin}  \& {Jackson}}{{Kaviraj} et~al.}{2025}]{Kaviraj2025}
{Kaviraj} S.,  {Lazar} I.,  {Watkins} A.~E.,  {Laigle} C.,  {Martin} G.,   {Jackson} R.~A.,  2025, \mn@doi [arXiv e-prints] {10.48550/arXiv.2502.02656}, \href {https://ui.adsabs.harvard.edu/abs/2025arXiv250202656K} {p. arXiv:2502.02656}

\bibitem[\protect\citeauthoryear{{Kewley}, {Dopita}, {Sutherland}, {Heisler}  \& {Trevena}}{{Kewley} et~al.}{2001}]{Kewley2001}
{Kewley} L.~J.,  {Dopita} M.~A.,  {Sutherland} R.~S.,  {Heisler} C.~A.,   {Trevena} J.,  2001, \mn@doi [\apj] {10.1086/321545}, \href {https://ui.adsabs.harvard.edu/abs/2001ApJ...556..121K} {556, 121}

\bibitem[\protect\citeauthoryear{{Kewley}, {Groves}, {Kauffmann}  \& {Heckman}}{{Kewley} et~al.}{2006}]{Kewley2006}
{Kewley} L.~J.,  {Groves} B.,  {Kauffmann} G.,   {Heckman} T.,  2006, \mn@doi [\mnras] {10.1111/j.1365-2966.2006.10859.x}, \href {https://ui.adsabs.harvard.edu/abs/2006MNRAS.372..961K} {372, 961}

\bibitem[\protect\citeauthoryear{{Khostovan} et~al.,}{{Khostovan} et~al.}{2025}]{Khostovan2025}
{Khostovan} A.~A.,  et~al., 2025, \mn@doi [arXiv e-prints] {10.48550/arXiv.2503.00120}, \href {https://ui.adsabs.harvard.edu/abs/2025arXiv250300120K} {p. arXiv:2503.00120}

\bibitem[\protect\citeauthoryear{{Kimura}, {Yamada}, {Kokubo}, {Yasuda}, {Morokuma}, {Nagao}  \& {Matsuoka}}{{Kimura} et~al.}{2020}]{Kimura2020}
{Kimura} Y.,  {Yamada} T.,  {Kokubo} M.,  {Yasuda} N.,  {Morokuma} T.,  {Nagao} T.,   {Matsuoka} Y.,  2020, \mn@doi [\apj] {10.3847/1538-4357/ab83f3}, \href {https://ui.adsabs.harvard.edu/abs/2020ApJ...894...24K} {894, 24}

\bibitem[\protect\citeauthoryear{{King} \& {Nealon}}{{King} \& {Nealon}}{2021}]{King2021}
{King} A.,  {Nealon} R.,  2021, \mn@doi [\mnras] {10.1093/mnrasl/slaa200}, \href {https://ui.adsabs.harvard.edu/abs/2021MNRAS.502L...1K} {502, L1}

\bibitem[\protect\citeauthoryear{{Koekemoer} et~al.,}{{Koekemoer} et~al.}{2007}]{Koekemoer2007}
{Koekemoer} A.~M.,  et~al., 2007, \mn@doi [\apjs] {10.1086/520086}, \href {https://ui.adsabs.harvard.edu/abs/2007ApJS..172..196K} {172, 196}

\bibitem[\protect\citeauthoryear{{Kormendy} \& {Richstone}}{{Kormendy} \& {Richstone}}{1995}]{Kormendy1995}
{Kormendy} J.,  {Richstone} D.,  1995, \mn@doi [\araa] {10.1146/annurev.aa.33.090195.003053}, \href {https://ui.adsabs.harvard.edu/abs/1995ARA&A..33..581K} {33, 581}

\bibitem[\protect\citeauthoryear{{Koudmani}, {Sijacki}, {Bourne}  \& {Smith}}{{Koudmani} et~al.}{2019}]{Koudmani2019}
{Koudmani} S.,  {Sijacki} D.,  {Bourne} M.~A.,   {Smith} M.~C.,  2019, \mn@doi [\mnras] {10.1093/mnras/stz097}, \href {http://adsabs.harvard.edu/abs/2019MNRAS.484.2047K} {484, 2047}

\bibitem[\protect\citeauthoryear{{Koudmani}, {Sijacki}  \& {Smith}}{{Koudmani} et~al.}{2022}]{Koudmani2022}
{Koudmani} S.,  {Sijacki} D.,   {Smith} M.~C.,  2022, \mn@doi [\mnras] {10.1093/mnras/stac2252}, \href {https://ui.adsabs.harvard.edu/abs/2022MNRAS.516.2112K} {516, 2112}

\bibitem[\protect\citeauthoryear{{Koudmani}, {Rennehan}, {Somerville}, {Hayward}, {Angl{\'e}s-Alc{\'a}zar}, {Orr}, {Sands}  \& {Wellons}}{{Koudmani} et~al.}{2024}]{Koudmani2024}
{Koudmani} S.,  {Rennehan} D.,  {Somerville} R.~S.,  {Hayward} C.~C.,  {Angl{\'e}s-Alc{\'a}zar} D.,  {Orr} M.~E.,  {Sands} I.~S.,   {Wellons} S.,  2024, \mn@doi [arXiv e-prints] {10.48550/arXiv.2409.02172}, \href {https://ui.adsabs.harvard.edu/abs/2024arXiv240902172K} {p. arXiv:2409.02172}

\bibitem[\protect\citeauthoryear{{Kristensen}, {Pimbblet}  \& {Penny}}{{Kristensen} et~al.}{2020}]{Kristensen2020}
{Kristensen} M.~T.,  {Pimbblet} K.,   {Penny} S.,  2020, \mn@doi [\mnras] {10.1093/mnras/staa1719}, \href {https://ui.adsabs.harvard.edu/abs/2020MNRAS.496.2577K} {496, 2577}

\bibitem[\protect\citeauthoryear{{Kuijken}}{{Kuijken}}{2011}]{Kuijken2011}
{Kuijken} K.,  2011, The Messenger, \href {https://ui.adsabs.harvard.edu/abs/2011Msngr.146....8K} {146, 8}

\bibitem[\protect\citeauthoryear{{Laigle} et~al.,}{{Laigle} et~al.}{2016}]{Laigle2016}
{Laigle} C.,  et~al., 2016, \mn@doi [\apjs] {10.3847/0067-0049/224/2/24}, \href {https://ui.adsabs.harvard.edu/abs/2016ApJS..224...24L} {224, 24}

\bibitem[\protect\citeauthoryear{{Laigle} et~al.,}{{Laigle} et~al.}{2018}]{Laigle2018}
{Laigle} C.,  et~al., 2018, \mn@doi [\mnras] {10.1093/mnras/stx3055}, \href {https://ui.adsabs.harvard.edu/abs/2018MNRAS.474.5437L} {474, 5437}

\bibitem[\protect\citeauthoryear{{Laureijs} et~al.,}{{Laureijs} et~al.}{2011}]{Laureijs2011}
{Laureijs} R.,  et~al., 2011, preprint, \href {http://adsabs.harvard.edu/abs/2011arXiv1110.3193L} {} (\mn@eprint {arXiv} {1110.3193})

\bibitem[\protect\citeauthoryear{{Lazar}, {Kaviraj}, {Martin}, {Laigle}, {Watkins}  \& {Jackson}}{{Lazar} et~al.}{2023}]{Lazar2023}
{Lazar} I.,  {Kaviraj} S.,  {Martin} G.,  {Laigle} C.,  {Watkins} A.,   {Jackson} R.~A.,  2023, \mn@doi [\mnras] {10.1093/mnras/stad224}, \href {https://ui.adsabs.harvard.edu/abs/2023MNRAS.520.2109L} {520, 2109}

\bibitem[\protect\citeauthoryear{{Lazar}, {Kaviraj}, {Watkins}, {Martin}, {Bichang'a}  \& {Jackson}}{{Lazar} et~al.}{2024a}]{Lazar2024a}
{Lazar} I.,  {Kaviraj} S.,  {Watkins} A.~E.,  {Martin} G.,  {Bichang'a} B.,   {Jackson} R.~A.,  2024a, \mn@doi [\mnras] {10.1093/mnras/stae510}, \href {https://ui.adsabs.harvard.edu/abs/2024MNRAS.529..499L} {529, 499}

\bibitem[\protect\citeauthoryear{{Lazar}, {Kaviraj}, {Watkins}, {Martin}, {Bichang'a}  \& {Jackson}}{{Lazar} et~al.}{2024b}]{Lazar2024b}
{Lazar} I.,  {Kaviraj} S.,  {Watkins} A.~E.,  {Martin} G.,  {Bichang'a} B.,   {Jackson} R.~A.,  2024b, \mn@doi [\mnras] {10.1093/mnras/stae1956}, \href {https://ui.adsabs.harvard.edu/abs/2024MNRAS.533.3771L} {533, 3771}

\bibitem[\protect\citeauthoryear{{Leauthaud} et~al.,}{{Leauthaud} et~al.}{2007}]{Leauthaud2007}
{Leauthaud} A.,  et~al., 2007, \mn@doi [\apjs] {10.1086/516598}, \href {https://ui.adsabs.harvard.edu/abs/2007ApJS..172..219L} {172, 219}

\bibitem[\protect\citeauthoryear{{Lintott} et~al.,}{{Lintott} et~al.}{2011}]{Lintott2011}
{Lintott} C.,  et~al., 2011, \mn@doi [\mnras] {10.1111/j.1365-2966.2010.17432.x}, \href {http://adsabs.harvard.edu/abs/2011MNRAS.410..166L} {410, 166}

\bibitem[\protect\citeauthoryear{{Lotz}, {Primack}  \& {Madau}}{{Lotz} et~al.}{2004}]{Lotz2004}
{Lotz} J.~M.,  {Primack} J.,   {Madau} P.,  2004, \mn@doi [\aj] {10.1086/421849}, \href {http://adsabs.harvard.edu/abs/2004AJ....128..163L} {128, 163}

\bibitem[\protect\citeauthoryear{{Marleau}, {Clancy}, {Habas}  \& {Bianconi}}{{Marleau} et~al.}{2017}]{Marleau2017}
{Marleau} F.~R.,  {Clancy} D.,  {Habas} R.,   {Bianconi} M.,  2017, \mn@doi [\aap] {10.1051/0004-6361/201629832}, \href {http://adsabs.harvard.edu/abs/2017A%26A...602A..28M} {602, A28}

\bibitem[\protect\citeauthoryear{{Martin} et~al.,}{{Martin} et~al.}{2019}]{Martin2019}
{Martin} G.,  et~al., 2019, \mn@doi [\mnras] {10.1093/mnras/stz356}, \href {http://adsabs.harvard.edu/abs/2019MNRAS.485..796M} {485, 796}

\bibitem[\protect\citeauthoryear{{Massey}, {Stoughton}, {Leauthaud}, {Rhodes}, {Koekemoer}, {Ellis}  \& {Shaghoulian}}{{Massey} et~al.}{2010}]{Massey2010}
{Massey} R.,  {Stoughton} C.,  {Leauthaud} A.,  {Rhodes} J.,  {Koekemoer} A.,  {Ellis} R.,   {Shaghoulian} E.,  2010, \mn@doi [\mnras] {10.1111/j.1365-2966.2009.15638.x}, \href {https://ui.adsabs.harvard.edu/abs/2010MNRAS.401..371M} {401, 371}

\bibitem[\protect\citeauthoryear{{McCracken} et~al.,}{{McCracken} et~al.}{2012}]{McCracken2012}
{McCracken} H.~J.,  et~al., 2012, \mn@doi [\aap] {10.1051/0004-6361/201219507}, \href {https://ui.adsabs.harvard.edu/abs/2012A&A...544A.156M} {544, A156}

\bibitem[\protect\citeauthoryear{{Mezcua} \& {Dom{\'\i}nguez S{\'a}nchez}}{{Mezcua} \& {Dom{\'\i}nguez S{\'a}nchez}}{2024}]{Mezcua2024}
{Mezcua} M.,  {Dom{\'\i}nguez S{\'a}nchez} H.,  2024, \mn@doi [\mnras] {10.1093/mnras/stae292}, \href {https://ui.adsabs.harvard.edu/abs/2024MNRAS.528.5252M} {528, 5252}

\bibitem[\protect\citeauthoryear{{Mezcua}, {Civano}, {Marchesi}, {Suh}, {Fabbiano}  \& {Volonteri}}{{Mezcua} et~al.}{2018}]{Mezcua2018}
{Mezcua} M.,  {Civano} F.,  {Marchesi} S.,  {Suh} H.,  {Fabbiano} G.,   {Volonteri} M.,  2018, \mn@doi [\mnras] {10.1093/mnras/sty1163}, \href {https://ui.adsabs.harvard.edu/abs/2018MNRAS.478.2576M} {478, 2576}

\bibitem[\protect\citeauthoryear{{Mezcua}, {Suh}  \& {Civano}}{{Mezcua} et~al.}{2019}]{Mezcua2019}
{Mezcua} M.,  {Suh} H.,   {Civano} F.,  2019, \mn@doi [\mnras] {10.1093/mnras/stz1760}, \href {https://ui.adsabs.harvard.edu/abs/2019MNRAS.488..685M} {488, 685}

\bibitem[\protect\citeauthoryear{{Mezcua}, {Pacucci}, {Suh}, {Siudek}  \& {Natarajan}}{{Mezcua} et~al.}{2024}]{Mezcua2024b}
{Mezcua} M.,  {Pacucci} F.,  {Suh} H.,  {Siudek} M.,   {Natarajan} P.,  2024, \mn@doi [\apjl] {10.3847/2041-8213/ad3c2a}, \href {https://ui.adsabs.harvard.edu/abs/2024ApJ...966L..30M} {966, L30}

\bibitem[\protect\citeauthoryear{{Moran}, {Shahinyan}, {Sugarman}, {V{\'e}lez}  \& {Eracleous}}{{Moran} et~al.}{2014}]{Moran2014}
{Moran} E.~C.,  {Shahinyan} K.,  {Sugarman} H.~R.,  {V{\'e}lez} D.~O.,   {Eracleous} M.,  2014, \mn@doi [\aj] {10.1088/0004-6256/148/6/136}, \href {http://adsabs.harvard.edu/abs/2014AJ....148..136M} {148, 136}

\bibitem[\protect\citeauthoryear{{Nyland} et~al.,}{{Nyland} et~al.}{2017}]{Nyland2017}
{Nyland} K.,  et~al., 2017, \mn@doi [\apj] {10.3847/1538-4357/aa7ecf}, \href {https://ui.adsabs.harvard.edu/abs/2017ApJ...845...50N} {845, 50}

\bibitem[\protect\citeauthoryear{{Papovich} et~al.,}{{Papovich} et~al.}{2018}]{Papovich2018}
{Papovich} C.,  et~al., 2018, \mn@doi [\apj] {10.3847/1538-4357/aaa766}, \href {https://ui.adsabs.harvard.edu/abs/2018ApJ...854...30P} {854, 30}

\bibitem[\protect\citeauthoryear{{Pardo} et~al.,}{{Pardo} et~al.}{2016}]{Pardo2016}
{Pardo} K.,  et~al., 2016, \mn@doi [\apj] {10.3847/0004-637X/831/2/203}, \href {https://ui.adsabs.harvard.edu/abs/2016ApJ...831..203P} {831, 203}

\bibitem[\protect\citeauthoryear{{Piotrowska}, {Bluck}, {Maiolino}  \& {Peng}}{{Piotrowska} et~al.}{2022}]{Piotrowska2022}
{Piotrowska} J.~M.,  {Bluck} A. F.~L.,  {Maiolino} R.,   {Peng} Y.,  2022, \mn@doi [\mnras] {10.1093/mnras/stab3673}, \href {https://ui.adsabs.harvard.edu/abs/2022MNRAS.512.1052P} {512, 1052}

\bibitem[\protect\citeauthoryear{{Pucha} et~al.,}{{Pucha} et~al.}{2025}]{Pucha2025}
{Pucha} R.,  et~al., 2025, \mn@doi [\apj] {10.3847/1538-4357/adb1dd}, \href {https://ui.adsabs.harvard.edu/abs/2025ApJ...982...10P} {982, 10}

\bibitem[\protect\citeauthoryear{{Reines}, {Greene}  \& {Geha}}{{Reines} et~al.}{2013}]{Reines2013}
{Reines} A.~E.,  {Greene} J.~E.,   {Geha} M.,  2013, \mn@doi [\apj] {10.1088/0004-637X/775/2/116}, \href {http://adsabs.harvard.edu/abs/2013ApJ...775..116R} {775, 116}

\bibitem[\protect\citeauthoryear{{Richstone} et~al.,}{{Richstone} et~al.}{1998}]{Richstone1998}
{Richstone} D.,  et~al., 1998, \mn@doi [\nat] {10.48550/arXiv.astro-ph/9810378}, \href {https://ui.adsabs.harvard.edu/abs/1998Natur.395A..14R} {385, A14}

\bibitem[\protect\citeauthoryear{{Sacchi}, {Bogd{\'a}n}, {Chadayammuri}  \& {Ricarte}}{{Sacchi} et~al.}{2024}]{Sacchi2024}
{Sacchi} A.,  {Bogd{\'a}n} {\'A}.,  {Chadayammuri} U.,   {Ricarte} A.,  2024, \mn@doi [\apj] {10.3847/1538-4357/ad684e}, \href {https://ui.adsabs.harvard.edu/abs/2024ApJ...974...14S} {974, 14}

\bibitem[\protect\citeauthoryear{{Satyapal}, {Abel}  \& {Secrest}}{{Satyapal} et~al.}{2018}]{Satyapal2018}
{Satyapal} S.,  {Abel} N.~P.,   {Secrest} N.~J.,  2018, \mn@doi [\apj] {10.3847/1538-4357/aab7f8}, \href {http://adsabs.harvard.edu/abs/2018ApJ...858...38S} {858, 38}

\bibitem[\protect\citeauthoryear{{Sawicki} et~al.,}{{Sawicki} et~al.}{2019}]{Sawicki2019}
{Sawicki} M.,  et~al., 2019, \mn@doi [\mnras] {10.1093/mnras/stz2522}, \href {https://ui.adsabs.harvard.edu/abs/2019MNRAS.489.5202S} {489, 5202}

\bibitem[\protect\citeauthoryear{{Schaap} \& {van de Weygaert}}{{Schaap} \& {van de Weygaert}}{2000}]{Schaap2000}
{Schaap} W.~E.,  {van de Weygaert} R.,  2000, \mn@doi [\aap] {10.48550/arXiv.astro-ph/0011007}, \href {https://ui.adsabs.harvard.edu/abs/2000A&A...363L..29S} {363, L29}

\bibitem[\protect\citeauthoryear{{Scholtz} et~al.,}{{Scholtz} et~al.}{2023}]{Scholtz2023}
{Scholtz} J.,  et~al., 2023, \mn@doi [arXiv e-prints] {10.48550/arXiv.2311.18731}, \href {https://ui.adsabs.harvard.edu/abs/2023arXiv231118731S} {p. arXiv:2311.18731}

\bibitem[\protect\citeauthoryear{{Schutte}, {Reines}  \& {Greene}}{{Schutte} et~al.}{2019}]{Schutte2019}
{Schutte} Z.,  {Reines} A.~E.,   {Greene} J.~E.,  2019, \mn@doi [\apj] {10.3847/1538-4357/ab35dd}, \href {https://ui.adsabs.harvard.edu/abs/2019ApJ...887..245S} {887, 245}

\bibitem[\protect\citeauthoryear{{Sharma}, {Brooks}, {Tremmel}, {Bellovary}, {Ricarte}  \& {Quinn}}{{Sharma} et~al.}{2022}]{Sharma2022}
{Sharma} R.~S.,  {Brooks} A.~M.,  {Tremmel} M.,  {Bellovary} J.,  {Ricarte} A.,   {Quinn} T.~R.,  2022, \mn@doi [\apj] {10.3847/1538-4357/ac8664}, \href {https://ui.adsabs.harvard.edu/abs/2022ApJ...936...82S} {936, 82}

\bibitem[\protect\citeauthoryear{{Sharma}, {Brooks}, {Tremmel}, {Bellovary}  \& {Quinn}}{{Sharma} et~al.}{2023}]{Sharma2023}
{Sharma} R.~S.,  {Brooks} A.~M.,  {Tremmel} M.,  {Bellovary} J.,   {Quinn} T.~R.,  2023, \mn@doi [\apj] {10.3847/1538-4357/ace046}, \href {https://ui.adsabs.harvard.edu/abs/2023ApJ...957...16S} {957, 16}

\bibitem[\protect\citeauthoryear{{Shimwell} et~al.,}{{Shimwell} et~al.}{2017}]{Shimwell2017}
{Shimwell} T.~W.,  et~al., 2017, \mn@doi [\aap] {10.1051/0004-6361/201629313}, \href {https://ui.adsabs.harvard.edu/abs/2017A&A...598A.104S} {598, A104}

\bibitem[\protect\citeauthoryear{{Silk} \& {Rees}}{{Silk} \& {Rees}}{1998}]{Silk1998}
{Silk} J.,  {Rees} M.~J.,  1998, \aap, \href {https://ui.adsabs.harvard.edu/abs/1998A&A...331L...1S} {331, L1}

\bibitem[\protect\citeauthoryear{{Siudek}, {Mezcua}  \& {Krywult}}{{Siudek} et~al.}{2023}]{Siudek2023}
{Siudek} M.,  {Mezcua} M.,   {Krywult} J.,  2023, \mn@doi [\mnras] {10.1093/mnras/stac3092}, \href {https://ui.adsabs.harvard.edu/abs/2023MNRAS.518..724S} {518, 724}

\bibitem[\protect\citeauthoryear{{Siudek} et~al.,}{{Siudek} et~al.}{2024}]{Siudek2024}
{Siudek} M.,  et~al., 2024, \mn@doi [\aap] {10.1051/0004-6361/202451761}, \href {https://ui.adsabs.harvard.edu/abs/2024A&A...691A.308S} {691, A308}

\bibitem[\protect\citeauthoryear{{Sousbie}}{{Sousbie}}{2011}]{Sousbie2011}
{Sousbie} T.,  2011, \mn@doi [\mnras] {10.1111/j.1365-2966.2011.18394.x}, \href {https://ui.adsabs.harvard.edu/abs/2011MNRAS.414..350S} {414, 350}

\bibitem[\protect\citeauthoryear{{Steinhardt} et~al.,}{{Steinhardt} et~al.}{2014}]{Steinhardt2014}
{Steinhardt} C.~L.,  et~al., 2014, \mn@doi [\apjl] {10.1088/2041-8205/791/2/L25}, \href {https://ui.adsabs.harvard.edu/abs/2014ApJ...791L..25S} {791, L25}

\bibitem[\protect\citeauthoryear{{Strateva} et~al.,}{{Strateva} et~al.}{2001}]{Strateva2001}
{Strateva} I.,  et~al., 2001, \mn@doi [\aj] {10.1086/323301}, \href {http://adsabs.harvard.edu/abs/2001AJ....122.1861S} {122, 1861}

\bibitem[\protect\citeauthoryear{{Taniguchi} et~al.,}{{Taniguchi} et~al.}{2007}]{Taniguchi2007}
{Taniguchi} Y.,  et~al., 2007, \mn@doi [\apjs] {10.1086/516596}, \href {https://ui.adsabs.harvard.edu/abs/2007ApJS..172....9T} {172, 9}

\bibitem[\protect\citeauthoryear{{Taniguchi} et~al.,}{{Taniguchi} et~al.}{2015}]{Taniguchi2015}
{Taniguchi} Y.,  et~al., 2015, \mn@doi [\pasj] {10.1093/pasj/psv106}, \href {https://ui.adsabs.harvard.edu/abs/2015PASJ...67..104T} {67, 104}

\bibitem[\protect\citeauthoryear{{Tolstoy}, {Hill}  \& {Tosi}}{{Tolstoy} et~al.}{2009}]{Tolstoy2009}
{Tolstoy} E.,  {Hill} V.,   {Tosi} M.,  2009, \mn@doi [\araa] {10.1146/annurev-astro-082708-101650}, \href {https://ui.adsabs.harvard.edu/abs/2009ARA&A..47..371T} {47, 371}

\bibitem[\protect\citeauthoryear{{Trebitsch}, {Volonteri}, {Dubois}  \& {Madau}}{{Trebitsch} et~al.}{2018}]{Trebitsch2018}
{Trebitsch} M.,  {Volonteri} M.,  {Dubois} Y.,   {Madau} P.,  2018, \mn@doi [\mnras] {10.1093/mnras/sty1406}, \href {https://ui.adsabs.harvard.edu/abs/2018MNRAS.478.5607T} {478, 5607}

\bibitem[\protect\citeauthoryear{{Trujillo} et~al.,}{{Trujillo} et~al.}{2021}]{Trujillo2021}
{Trujillo} I.,  et~al., 2021, \mn@doi [\aap] {10.1051/0004-6361/202141603}, \href {https://ui.adsabs.harvard.edu/abs/2021A&A...654A..40T} {654, A40}

\bibitem[\protect\citeauthoryear{{Venhola} et~al.,}{{Venhola} et~al.}{2018}]{Venhola2018}
{Venhola} A.,  et~al., 2018, \mn@doi [\aap] {10.1051/0004-6361/201833933}, \href {https://ui.adsabs.harvard.edu/abs/2018A&A...620A.165V} {620, A165}

\bibitem[\protect\citeauthoryear{{Ward}, {Harrison}, {Costa}  \& {Mainieri}}{{Ward} et~al.}{2022}]{Ward2022}
{Ward} S.~R.,  {Harrison} C.~M.,  {Costa} T.,   {Mainieri} V.,  2022, \mn@doi [\mnras] {10.1093/mnras/stac1219}, \href {https://ui.adsabs.harvard.edu/abs/2022MNRAS.514.2936W} {514, 2936}

\bibitem[\protect\citeauthoryear{{Watkins}, {Kaviraj}, {Collins}, {Knapen}, {Kelvin}, {Duc}, {Rom{\'a}n}  \& {Mihos}}{{Watkins} et~al.}{2024}]{Watkins2024}
{Watkins} A.~E.,  {Kaviraj} S.,  {Collins} C.~C.,  {Knapen} J.~H.,  {Kelvin} L.~S.,  {Duc} P.-A.,  {Rom{\'a}n} J.,   {Mihos} J.~C.,  2024, \mn@doi [\mnras] {10.1093/mnras/stae236}, \href {https://ui.adsabs.harvard.edu/abs/2024MNRAS.528.4289W} {528, 4289}

\bibitem[\protect\citeauthoryear{{Weaver} et~al.,}{{Weaver} et~al.}{2022}]{Weaver2022}
{Weaver} J.~R.,  et~al., 2022, \mn@doi [\apjs] {10.3847/1538-4365/ac3078}, \href {https://ui.adsabs.harvard.edu/abs/2022ApJS..258...11W} {258, 11}

\bibitem[\protect\citeauthoryear{{Whitaker}, {van Dokkum}, {Brammer}  \& {Franx}}{{Whitaker} et~al.}{2012}]{Whitaker2012}
{Whitaker} K.~E.,  {van Dokkum} P.~G.,  {Brammer} G.,   {Franx} M.,  2012, \mn@doi [\apj] {10.1088/2041-8205/754/2/L29}, \href {https://ui.adsabs.harvard.edu/#abs/2012ApJ...754L..29W} {754, L29}

\bibitem[\protect\citeauthoryear{{Witten} et~al.,}{{Witten} et~al.}{2025}]{Witten2025}
{Witten} C.,  et~al., 2025, \mn@doi [\mnras] {10.1093/mnras/staf001}, \href {https://ui.adsabs.harvard.edu/abs/2025MNRAS.537..112W} {537, 112}

\bibitem[\protect\citeauthoryear{{Wright} et~al.,}{{Wright} et~al.}{2017}]{Wright2017}
{Wright} A.~H.,  et~al., 2017, \mn@doi [\mnras] {10.1093/mnras/stx1149}, \href {https://ui.adsabs.harvard.edu/abs/2017MNRAS.470..283W} {470, 283}

\bibitem[\protect\citeauthoryear{{Zabludoff} et~al.,}{{Zabludoff} et~al.}{2021}]{Zabludoff2021}
{Zabludoff} A.,  et~al., 2021, \mn@doi [\ssr] {10.1007/s11214-021-00829-4}, \href {https://ui.adsabs.harvard.edu/abs/2021SSRv..217...54Z} {217, 54}

\bibitem[\protect\citeauthoryear{{Zamojski} et~al.,}{{Zamojski} et~al.}{2007}]{Zamojski2007}
{Zamojski} M.~A.,  et~al., 2007, \mn@doi [\apjs] {10.1086/516593}, \href {https://ui.adsabs.harvard.edu/abs/2007ApJS..172..468Z} {172, 468}

\bibitem[\protect\citeauthoryear{{Zou} et~al.,}{{Zou} et~al.}{2023}]{Zou2023}
{Zou} F.,  et~al., 2023, \mn@doi [\apj] {10.3847/1538-4357/acce39}, \href {https://ui.adsabs.harvard.edu/abs/2023ApJ...950..136Z} {950, 136}

\bibitem[\protect\citeauthoryear{{van der Marel}, {Alves}, {Hardy}  \& {Suntzeff}}{{van der Marel} et~al.}{2002}]{Vandermarel2002}
{van der Marel} R.~P.,  {Alves} D.~R.,  {Hardy} E.,   {Suntzeff} N.~B.,  2002, \mn@doi [\aj] {10.1086/343775}, \href {https://ui.adsabs.harvard.edu/abs/2002AJ....124.2639V} {124, 2639}

\makeatother
\end{thebibliography}


\appendix

\section{Comparison of spectroscopic and photometric redshifts}

\label{app:pzsz}

\begin{figure}
\center
\includegraphics[width=\columnwidth]{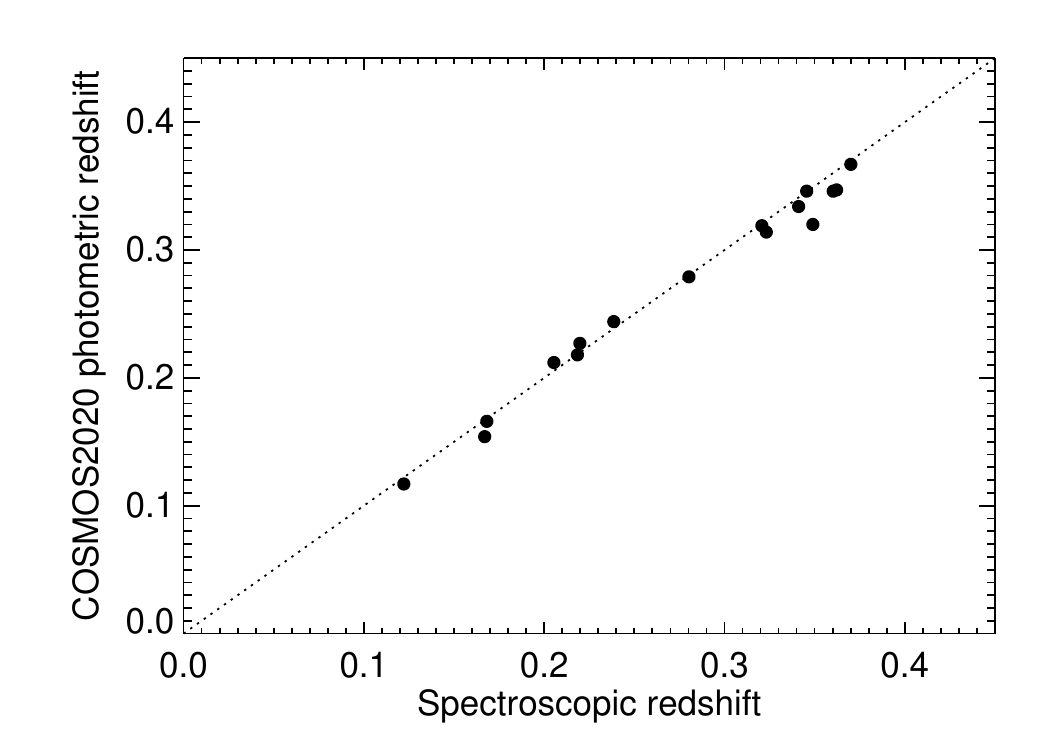}
\caption{{\color{black}Comparison of spectroscopic redshifts and photometric redshifts from COSMOS2020 for 16 variable dwarfs which have reliable spectroscopic redshifts from the compilation by \citet{Khostovan2025}.}} 
\label{fig:pzsz}
\end{figure}

{\color{black}Figure \ref{fig:pzsz} shows a comparison of spectroscopic redshifts ($z_{\rm spec}$) and photometric redshifts ($z_{\rm phot}$) of 16 variable dwarfs which have reliable spectroscopic redshifts (i.e. with confidence levels greater than 90 per cent) from the recent compilation by \citet{Khostovan2025}. This comprehensive catalogue contains 165,312 redshifts of 97,929 unique objects from 108 individual observing programmes. The photometric redshifts, which are derived via SED fitting, show excellent correspondence with their spectroscopic counterparts. The median difference between the photometric and spectroscopic redshifts is $\sim$0.0066. The median fractional difference, defined as ($z_{\rm spec}$ - $z_{\rm phot}$) / $z_{\rm spec}$ is 2.8 per cent.} 

\bsp	
\label{lastpage}
\end{document}